\documentclass[%
 aip,
 amsmath,amssymb,
 reprint,%
]{revtex4-2}
\usepackage{graphicx}
\usepackage{dcolumn}
\usepackage{bm}
\usepackage{xcolor}
\usepackage[utf8]{inputenc}
\usepackage[T1]{fontenc}
\usepackage{mathptmx}
\usepackage{etoolbox}
\usepackage{gensymb}
\usepackage{hyperref}

\makeatletter
\def\@email#1#2{%
 \endgroup
 \patchcmd{\titleblock@produce}
  {\frontmatter@RRAPformat}
  {\frontmatter@RRAPformat{\produce@RRAP{*#1\href{mailto:#2}{#2}}}\frontmatter@RRAPformat}
  {}{}
}%
\makeatother

\begin{document}

\preprint{AIP/123-QED}

\title{Origin of magnetic switching cascades in tetrahedral CoFe nanostructures}
\author{Christian Schröder}
\email{christian.schroeder@hsbi.de, christian.schroeder@uni-bielefeld.de}
\affiliation{Bielefeld Institute for Applied Materials Research, Bielefeld University of Applied Science and Arts, 33619 Bielefeld, Germany}
\affiliation{Faculty of Physics, Bielefeld University, 33501 Bielefeld, Germany}
\author{Bereket Ghebretinsae}
\affiliation{Institute of Physics, Goethe-University Frankfurt, 60438 Frankfurt am Main, Germany}
\author{Martin Lonsky}
\affiliation{Institute of Physics, Goethe-University Frankfurt, 60438 Frankfurt am Main, Germany}
\author{Mohanad Al Mamoori}
\affiliation{Institute of Physics, Goethe-University Frankfurt, 60438 Frankfurt am Main, Germany}
\author{Fabrizio Porrati}
\affiliation{Institute of Physics, Goethe-University Frankfurt, 60438 Frankfurt am Main, Germany}
\author{Michael Huth}
\affiliation{Institute of Physics, Goethe-University Frankfurt, 60438 Frankfurt am Main, Germany}
\author{Jens Müller}
\affiliation{Institute of Physics, Goethe-University Frankfurt, 60438 Frankfurt am Main, Germany}

\date{\today}

\begin{abstract}
We present a comprehensive study of small-scale three-dimensional (3D) tetrahedral CoFe nanostructure arrays prepared by focused electron beam-induced deposition (FEBID) and placed in two distinct orientations with respect to the direction of an external magnetic field. Using ultra-sensitive micro-Hall magnetometry we obtain angular-dependent magnetic stray field hysteresis loops that show characteristic cascading magnetic switching close to zero magnetic field. By employing micromagnetic simulations we could reproduce the hysteresis loops and identify characteristic field dependent magnetic configurations including a vortex-type groundstate. From this we derive a coarse-graining macrospin model and show that the complex switching behavior can be explained by the reorientation dynamics of non-interacting uniaxial anisotropic magnetic grains modeled as a superposition of Stoner-Wohlfarth particles. 
\end{abstract}

\maketitle
Three-dimensional (3D) magnetic nanostructures have become the subject of intense research both in fabrication of ever more complex structures and in applying sophisticated measuring techniques of both, large arrays or individual structures \cite{Fernandez-Pacheco2017,Streubel_2016,Donnelly2017,Sheka2022}. Lifting the 2D confinement of magnetic structures makes available additional degrees of freedom and leads to new fundamental effects such as topological spin textures \cite{Donnelly2022} or exotic dynamics \cite{Finizio2022} as well as the possibility for new applications, e.g.\ in high-density data storage \cite{Koraltan2021}.
The recent advancements in fabrication methods of 3D nanomagnets as e.g.\ in electrochemical deposition, two-photon lithography or 3D nanoprinting 
enable tailored structures with intriguing properties and create geometry- and curvature-induced effective interactions, see, e.g., \cite{Fernandez-Pacheco2017,Streubel_2021} and references therein. 
As building blocks of larger 3D architectures a wealth of different structures have been demonstrated, among them complex 3D tetrapod arrays \cite{Williams2018,Koraltan2021}, arrays of multi-axial nano-cubes and nano-trees \cite{Keller2018,AlMamoori2018}, simple devices such as a nanomagnetic conduits \cite{Sanz-2017} or, recently, curved structures like spirals or twisted ribbons \cite{Streubel_2016,Streubel_2021,Sheka2022}.  
One notable 3D printing method with high spatial resolution is focused electron beam-induced deposition (FEBID) which allows for the fabrication of high-purity magnetic deposits with tunable growth characteristics, see e.g.~\cite{Plank2020,Pacheco2020,Magen2021} for recent reviews. Conveniently, FEBID-grown magnetic nanostructures can be directly written on top of a micro-Hall sensor for local magnetic stray field measurements \cite{Porrati_2015,Pohlit2015,Pohlit2016b,Keller2018}. This versatile, ultra-sensitive method allows to investigate the switching characteristic of both large arrays and individual nano-scale magnetic structures \cite{Wirth2000,Li2002,Pohlit2016a,Pohlit2016b,AlMamoori2020}. 

\begin{figure}
	\centering
    \includegraphics[width=0.8\linewidth]{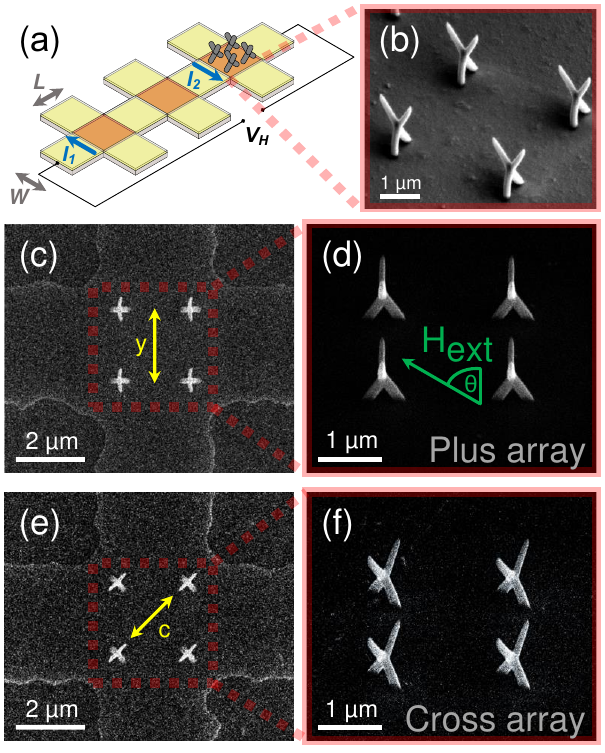}
	\caption{\label{SEM} Schematic of a micro-Hall sensor and SEM micrographs of CoFe tetrahedral nanostructures grown onto $3\times3$\,$\mu$m$^{2}$ Hall crosses by FEBID. (a) Gradiometry setup with electric currents $I_1$ and $I_2=-I_1$ applied to an empty reference cross and a cross decorated with magnetic structures, respectively. The background Hall signal thereby is subtracted \textit{in situ}. (b) Side view of a CoFe tetrapod array. (c) Top and (d) front view of a 2$\times$2 array placed in 'plus' arrangement. (e) Top and (f) front view of 2$\times$2 array placed in 'cross' arrangement. The closest two upper arms of the tetrapods face each other along the $y$- and $c$-axes for 'plus' and 'cross' arrays, respectively. For the angle $\Theta = 0$ the external field is applied perpendicular to the Hall sensor.}  
\end{figure}

In this Letter we present new insights into the magnetization switching dynamics of FEBID-grown small-scale arrays of 3D CoFe tetrahedral nanostructures placed in what we term as 'plus' and 'cross' configurations on top of a micro-Hall magnetometer. The structures differ in their relative orientation with respect to each other and the external magnetic field, see Fig.~\ref{SEM}. We obtain angular-dependent magnetic stray field hysteresis loops that exhibit a striking cascade switching characteristics upon crossing zero external magnetic field. We show that on the basis of micromagnetic simulations a coarse-graining macrospin model, combining low-field and high-field magnetization states, can be derived that explains these characteristic features accurately by the reorientation dynamics of non-interacting magnetic grains with uniaxial anisotropy.

The 3D magnetic nanostructure arrays were directly written onto the active area of a $3\times3$\,$\mu$m$^{2}$ GaAs/AlGaAs micro-Hall sensor by FEBID and consist of metallic bcc Co$_3$Fe. 
The elementary units are identical tetrahedral nanostructures placed in two small-scale lattice geometries as $2 \times 2$ arrays in 'plus' and 'cross' arrangements shown in Figure \ref{SEM}(b)-(f) as scanning electron microscope (SEM) images. One tetrapod's four pillars are $\sim 500$\,nm in length and $\sim 130$\,nm in diameter each. The perpendicular, $z$-component of the magnetic stray field emanating from the 3D nanostructures averaged over the active area of the Hall cross, $\langle {B_z} \rangle$, is determined by measuring the Hall voltage $V_H$ generated in the high-mobility two-dimensional electron gas at the interface of a GaAs/AlGaAs heterostructure. 
An \emph{\it in situ} cancellation of the large linear background Hall signal was conducted by a differential measurement of an empty reference cross in a so-called gradiometry setup, see Fig.~\ref{SEM}(a), such that $\Delta {V_H} = \frac{1}{{ne}} \cdot I \cdot \langle {B_z} \rangle $, where $I$ and $n$ are the applied current and the carrier density, respectively. The micro-Hall magnetometer was placed inside a cryogenic system and magnetic field loops have been measured at $T = 30$\,K with varying field angles with respect to the sensor plane, where the reference angle $\theta  = 0$\degree\ indicates an applied field perpendicular to the plane, cf.~Fig.~\ref{SEM}(d).

\begin{figure*}
	\includegraphics[width=0.8\textwidth]{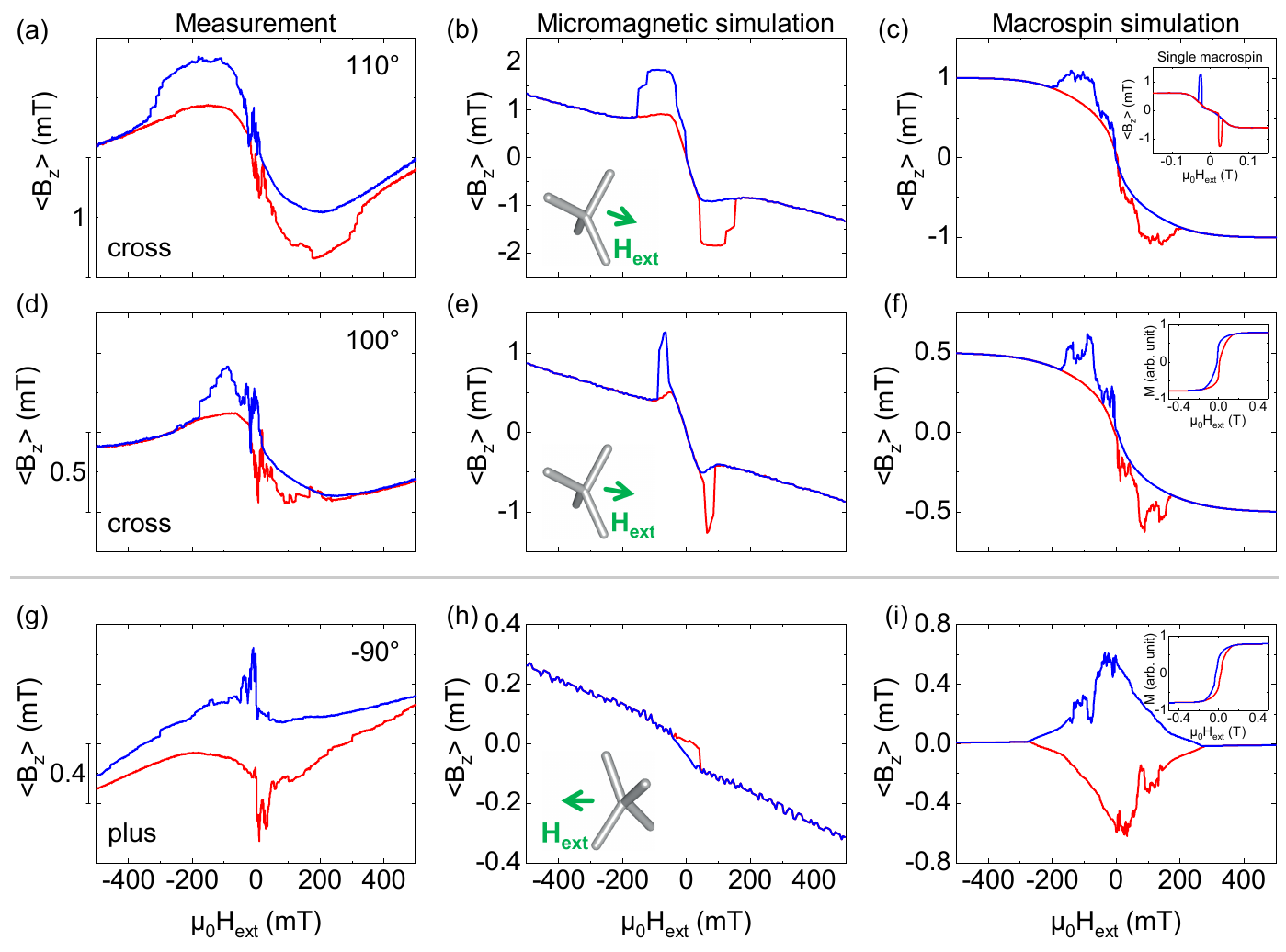}
	\caption{\label{ex_mm_ms_100deg} (a,d,g) Experimental stray-field hysteresis curve of the nanostructure array in 'cross' orientation measured at $\theta = 110\degree, 100$\degree and in 'plus' orientation at $\theta = -90$\degree exhibiting characteristic spike features around $H_{\rm ext} = 0$ in the up cycle (red) and point-symmetrically located in the down cycle (blue). (b,e,h) Micromagnetic and (c,f,i) macrospin simulations yielding hysteresis loops that reproduce these features. Surprisingly, a model with just one macrospin per pillar, inset of (c), already exhibits the characteristic overall shape of the hysteresis. Insets of (f,i) show the corresponding magnetization curves, which are not sensitive to the spike features.} 
\end{figure*}
Figures~\ref{ex_mm_ms_100deg}(a),(d),(g) exemplarily show the experimental stray field curves at angles of $\theta = 110\degree$ and $100\degree$ for the 'cross', and $-90\degree$ for the 'plus' array. Hysteresis loops for other angles are given in the Supplemental Information (SI) \cite{Supplement}. While the hysteresis loops of the {\it magnetization} show no unusual features besides the expected constriction at small fields, see inset in Fig.~\ref{ex_mm_ms_100deg}(f), for both structures we observe a peculiar overall shape in the {\it magnetic stray field} characteristic for the particular field angle. In addition, upon increasing the external field from negative saturation the curves initially are very smooth until $\mu_0H_{\mathrm{ext}} \approx 0\,$T. Upon further increasing the field and thus crossing zero, a series of pronounced spikes appear, see Fig.~\ref{ex_mm_ms_100deg}(a,d,g). The same behavior is found for the down cycle starting from positive saturation producing a hysteresis loop, which is nearly perfectly point symmetric in all details with respect to the origin. 
In order to understand the general shape and the particular spikes we have performed micromagnetic simulations taking into account previous results from microstructure analysis \cite{Keller2018}. On that basis we derive a coarse-graining macrospin model that not only reproduces the observed features but also yields the physical explanation.

\begin{figure}
	\includegraphics[width=0.9\linewidth]{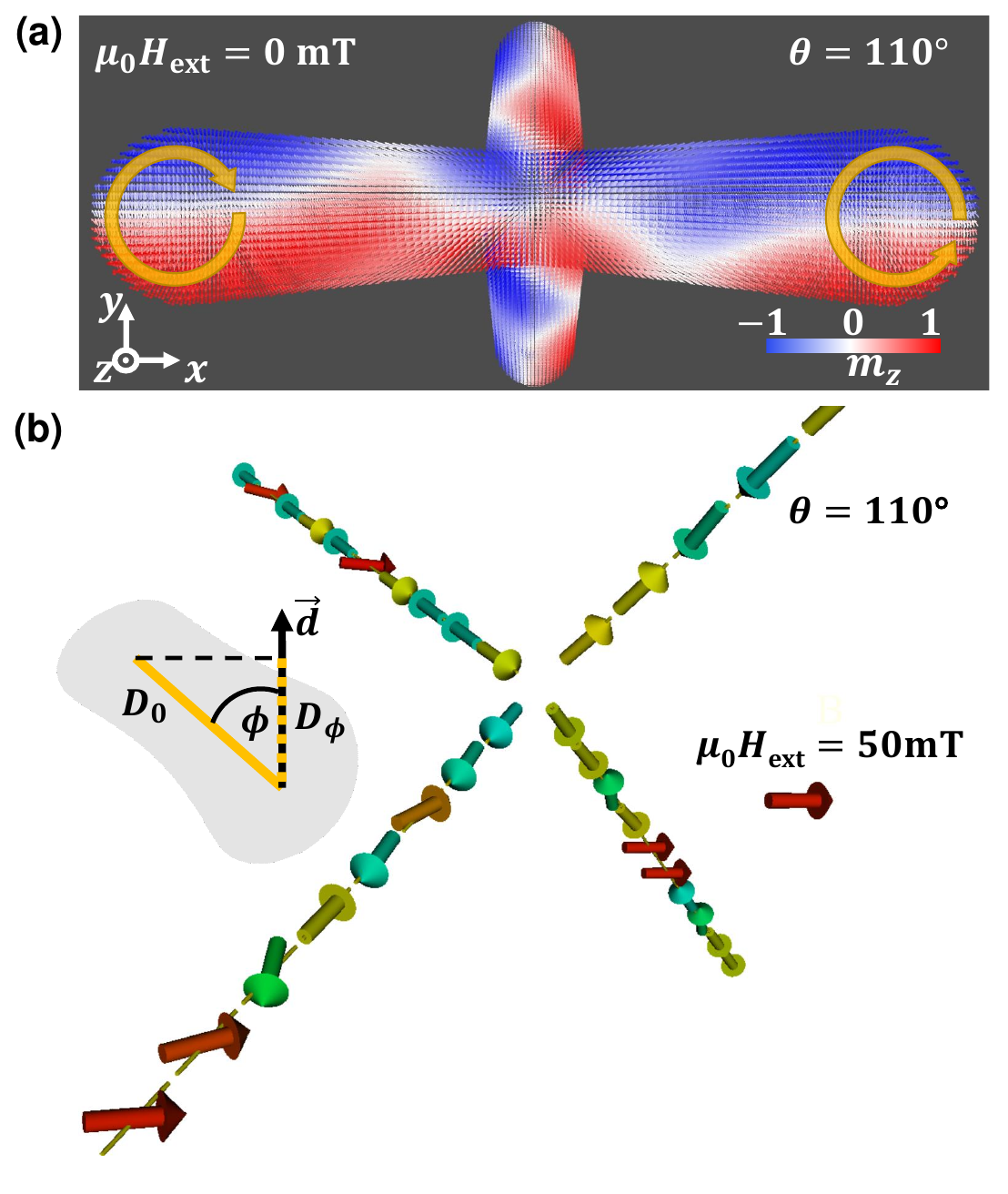}
	\caption{\label{vortex_state} (a) Snapshot of the magnetic configuration as obtained from micromagnetic simulations of a nano-tetrapod in the 'cross' array at field angle $\theta = 110$\degree\ and $\mu_0 H_{\mathrm{ext}}=0\,$mT. The magnetic configuration of each pillar collapses into a cylindrical vortex state. (b) Snapshot of the magnetic configuration as obtained from macrospin simulations at field angle $\theta = 110$\degree\ and $\mu_0 H_{\mathrm{ext}}=50\,$mT (red arrow on the right). Depending on the local coercive fields $H_{\mathrm{C},\phi}$ (anisotropies shown as yellow lines at each spin site) the macrospins abruptly change their orientations and follow the external field. For visualization purposes we show a model with just 10 macrospins per pillar. The inset on the left shows the schematic sketch of the magnetic grain model, i.e.\ the projection of a grain's anisotropy axis onto the direction of the pillar axis.} 
\end{figure}

Results of electron diffraction analysis indicate that the microstructure of each pillar is nano-granular and consists of randomly oriented bcc-structured Co-rich CoFe grains with sizes of about $5\,$nm \cite{Keller2018}. Assuming a random orientation of the grains we expect magneto-crystalline anisotropy effects canceling out due to averaging. Furthermore, we find an exponentially decreasing oxidation of the material from the surface of the pillars towards the inside which in turn leads to a decreasing saturation magnetization from inside towards the surface of the pillars. Here, we use a simplified micromagnetic model in which the saturation magnetization is suitably chosen in order to obtain a satisfying agreement with the experiment. 
We performed simulations using the \textit{mumax$^3$} software\cite{Vansteenkiste_2014} and the following parameters: saturation magnetization $M_{\mathrm{S}}=1.25\,$MA/m, exchange stiffness $A_{\mathrm{ex}}=17.25\,$pJ/m, Gilbert damping $\alpha=0.3$. Comparing the micromagnetic results shown in Fig.~\ref{ex_mm_ms_100deg}(b,e) with the measurements, Fig.~\ref{ex_mm_ms_100deg}(a,d), we find a reasonably good qualitative agreement with the overall shape of the hysteresis loop, namely the broad negative peak for positive fields on the up cycle and a broad positive peak for negative fields on the down cycle. Note, that the results of Fig.~\ref{ex_mm_ms_100deg}(h) significantly deviate from the experimental results which we will explain below. In Fig.~\ref{vortex_state}(a) we show a snapshot of the magnetic configuration of a 'cross' array nanostructure at field angle $\theta = 110$\degree\ and $\mu_0 H_{\mathrm{ext}}=0\,$mT. For fields at and close to zero, the magnetic configuration of each pillar collapses into cylindrical vortex states. Taking into account the length/diameter ratio of the pillars and negligible dipole-dipole interaction between the grains, vortex-like magnetization states are energetically favorable which is consistent with our micromagnetic results. At higher fields the magnetic configuration changes to a unaxial magnetization along the pillar axes with different ``in-out'' directions, and a further alignment of the magnetization with the external field occurs when approaching saturation. 

Starting from the uniaxial magnetic configurations at high fields we first derive the simplest possible macrospin model. Here, each pillar is modeled by just {\it one} macrospin with magnetic moment $\vec m_i(\vec r_i)$ placed in the center position of each pillar and with uniaxial anisotropy of strength $D_0$ directed along the pillar axes. The four macrospins of each tetrapod are non-interacting, i.e.\ only their relative orientation to the external field and placement with respect to the sensor plane is relevant for their stray field contribution. The stray fields are calculated according to 
\begin{equation}
\label{calc_bz}
   \langle {B_z} \rangle= \langle \sum_{i=1}^{N} \vec b_z(m_i(\vec r_i)) \rangle
\end{equation}
as a superposition of the $N=16$ individual stray fields $\vec b_z(m_i(\vec r_i))$ of the macrospins averaged over the active area of the micro-Hall sensor. A macrospin's dynamics can be solely described within the framework of the Stoner-Wohlfarth model \cite{Stoner1948}. We have calculated the stray field component $\langle B_z \rangle$ by means of spin dynamics simulations \cite{Keller2018} for all experimentally probed directions of the external magnetic field. According to the Stoner-Wohlfarth model, a macrospin has a coercive field of $H_{\mathrm{C}}=2D_0/M_{\mathrm{S}}$, where $M_{\mathrm{S}}$ is the saturation magnetization. By adjusting $D_0$ and $M_{\mathrm{S}}$ in the simulations we can scale our hysteresis curves to find best possible correspondence with the experiment. In the inset of Fig.~\ref{ex_mm_ms_100deg}(c) we show results for $\theta=110$\degree. Even though this is a crudely oversimplified model it reflects the overall shape found in the experiment as well as in the micromagnetic simulations, namely a negative peak at $H_{\mathrm{C}}$ on the up cycle and and a positive peak at $-H_{\mathrm{C}}$ on the down cycle. 

We now refine our macrospin model so that it also matches the low-field vortex-like magnetic configurations of the pillars where the magnetization of each voxel is oriented along their curved surfaces. Topologically, this magnetic configuration is equivalent to the one where the magnetization directions of all voxels point radially outward since the range of orientation angles for a radial and a chiral spin orientation is identical, except in the vortex core. From this we can derive an effective coarse-graining macrospin model in the following way. We now assume that each grain within a pillar is fully characterized by its total magnetic moment and uniaxial anisotropy with strength $D_0$. Since the grains are randomly oriented, their anisotropy axes are randomly oriented as well. This is equivalent to the situation where all grain anisotropies point along one direction $\vec d$ but have different effective anisotropy strengths according to the projection onto $\vec d$, see inset of Fig.~\ref{vortex_state}(b). If $\phi$ is the polar angle of a grain's anisotropy axis with respect to $\vec d$, then $D_\phi = D_0 \cdot \cos\phi$ is the effective anisotropy strength. For randomly oriented grains $\phi$ is evenly random distributed in the interval $[0;\pi/2]$. However, $D_\phi$ is not evenly distributed in the interval $[D_0;0]$ but peaked at small angles. Thus, there is a bias for larger anisotropies in the direction of the pillar axes. However, in our model we want to realize the opposite effect, i.e. the preferred magnetization direction is pointing radially outward, i.e.~perpendicular to the pillar axis. Therefore, we use the inverted distribution given by $D_\phi = D_0 \cdot \left(1-\cos\phi\right)$. We have chosen each pillar to consist of 50 magnetic grains. In the Stoner-Wohlfarth model each according macrospin has a coercive field of $H_{\mathrm{C}, \phi}=2D_\phi/M_{\mathrm{S}}$ and thus, $H_{\mathrm{C},\phi}$ becomes a switching field distribution.  
In Fig.~\ref{ex_mm_ms_100deg}(c,f,i) we show results for our coarse-graining model at the given angles for the 'cross' and 'plus' structures using $N=(50 \times 4)\times 4=800$ macrospins. The qualitative agreement of the macrospin simulations with the experimental and micromagnetic results is striking, in particular regarding the observed steps and spikes. The overall shape of the curves agrees very well with the micromagnetic simulations for $\theta \neq \pm 90^\circ$.
Note that the macrospin model curves are strictly point-symmetric with respect to the origin which is a result of assuming \textit{non-interacting} Stoner-Wohlfarth particles. To a large extent our experimental curves show point symmetry as well which corroborates our model considerations. 
The good agreement of the experimental and macrospin results seems surprising and can be explained as follows. Starting from positive saturation, the macrospins point in the direction of the external field. Upon lowering the field until $\mu_0H_{\mathrm{ext}}=0\,$T the macrospins continuously rotate and relax towards their ground state, i.e.\ an alignment with their \textit{local} anisotropy axes. However, after crossing $\mu_0H_{\mathrm{ext}}=0\,$T and depending on the angle of the anisotropy axes with respect to the external field, the macrospins abruptly change their orientation at fields $H_{\mathrm{C},\phi}$ \cite{Tannous2008}, see Fig.~\ref{vortex_state}(b), leading to a switching cascade and hence the observed spikes and steps in the stray field measurements shown in Fig.~\ref{ex_mm_ms_100deg}. 

Note that the spikes around $\mu_0H_{\mathrm{ext}}=0\,$T do not appear in our micromagnetic simulations. This is due to methodological reasons as we are required to use small cell sizes of edge length 5\,nm for the finite difference discretization in order to maintain the necessary accuracy. At around $\mu_0H_{\mathrm{ext}}=0\,$T the pillars are in cylindrical vortex states, thus the magnetization is gradually changing over each cell which has two effects: (i) The total magnetization of the structure is rather small due to canceling effects leading to a small total stray field. (ii) Small changes in $H_\mathrm{ext}$ only lead to small gradual changes of the magnetization per cell and thus cause only small stray field changes, i.e.\ no spikes. In contrast to this, each macrospin represents a large magnetic moment that switches at $H_\mathrm{C}$ resulting in large stray field changes and spikes, even at fields around $\mu_0H_{\mathrm{ext}}=0\,$T.
 
As a prediction of our coarse-graining model one should find a characteristic $\theta$-dependence of the total stray field amplitude $\Delta B_z(\theta)  = \langle B_z^\mathrm{max} \rangle - \langle B_z^\mathrm{min} \rangle$ for each hysteresis loop. For $\theta = \pm 90$\degree\ the external field is exactly parallel to the Hall cross, i.e.~points along the $x$-direction. At saturation and on lowering the field the grains' magnetization is more or less pointing in the $x$-direction as well. Thus, the $z$-component of the grains' magnetization is close to zero and their stray field contribution to $\langle B_z \rangle$ is very small for most fields, except for the fields $H_{\mathrm{C},\phi}$ at which the grains flip. The hysteresis curves at and around $\theta = \pm 90$\degree\ therefore show the most noticeable ``fingerprint'' of the grain switching cascade, however resulting in the smallest total stray field amplitude $\Delta B_z(\theta)$. On decreasing $\theta$, the $z$-component of the grains' magnetization increases and thus $\Delta B_z(\theta)$ increases as well. As a consequence, the relative stray field contribution of the grain switching cascade to $\Delta B_z(\theta)$ becomes smaller, leading to less pronounced spikes and steps with the least contribution at $\theta = 0$\degree. Indeed, we find this behaviour in the experimental and micromagnetic data, see the SI for details\cite{Supplement}.\\        
%
For a further proof of our model we have performed experiments with field sweeps where $\mu_0H_{\mathrm{ext}}=0\,$T is crossed and field sweeps where $\mu_0H_{\mathrm{ext}}=0\,$T is not crossed. In the latter case one expects fewer spikes and steps since the coercive fields $H_{\mathrm{C},\phi}$ can not be reached and thus grain flips do not occur. Indeed, we observe that avoiding zero-field crossing while sweeping the field leads to fewer and smaller spikes and steps, see SI\cite{Supplement}.  


In conclusion, we have successfully modeled the magnetization reversal of FEBID-prepared small-scale arrays of tetrahedral CoFe nanostructures placed in 'plus' and 'cross' configurations with respect to the direction of an external magnetic field. We have shown that the physics of the characteristic features of the measured stray field hysteresis loops can be understood by considering non-interacting magnetic grains, each described by a simple Stoner-Wohlfarth model with anisotropy axis aligned along the pillar directions of the nanostructures. By assuming a distribution of anisotropy strengths due to a random distribution of the grain orientations we obtain a distribution of coercive fields. This leads to magnetic switching cascades of the grains during a field sweep which manifest themselves as steps and spikes in stray field measurements. The coarse-graining macrospin simulations catch the essential switching dynamics of vortex-like magnetization states with rather short computing times of a few minutes per angle on a laptop as compared to several hours for GPU-accelerated micromagnetic simulations. This may be advantageous when calculating larger and more complex structures.



\subsection*{Acknowledgements} The authors thank Prof.~Dr.~Jürgen Weis from the Max-Planck Institute for Solid State Research (Stuttgart, Germany) for providing the GaAs/AlGaAs high-mobility wafer material. M.L.~acknowledges financial support through the FOKUS program by the Goethe Research Academy for Early Career Researchers (GRADE). The contribution of M.H. was supported by DFG funding (project HU 752/16-1; Frankfurt Center for Electron Microscopy (FCEM)).
\section*{Author Declarations}
The authors have no conflicts to disclose.
\section*{Data availability}
The data that support the findings of this study are available from the corresponding author upon reasonable request.

\bibliography{bibliography}

\end{document}



\title{Supplemental information for "Origin of magnetic switching cascades in tetrahedral CoFe nanostructures"}

\author{Christian Schröder}
\email{christian.schroeder@hsbi.de, christian.schroeder@uni-bielefeld.de}
\affiliation{Institute for Applied Materials Research, Bielefeld University of Applied Science and Arts, 33619 Bielefeld, Germany}
\affiliation{Faculty of Physics, Bielefeld University, 33501 Bielefeld, Germany}
\author{Bereket Ghebretinsae}
\affiliation{Institute of Physics, Goethe-University Frankfurt, 60438 Frankfurt am Main, Germany}
\author{Martin Lonsky}
\affiliation{Institute of Physics, Goethe-University Frankfurt, 60438 Frankfurt am Main, Germany}
\author{Mohanad Al Mamoori}
\affiliation{Institute of Physics, Goethe-University Frankfurt, 60438 Frankfurt am Main, Germany}
\author{Fabrizio Porrati}
\affiliation{Institute of Physics, Goethe-University Frankfurt, 60438 Frankfurt am Main, Germany}
\author{Michael Huth}
\affiliation{Institute of Physics, Goethe-University Frankfurt, 60438 Frankfurt am Main, Germany}
\author{Jens Müller}
\affiliation{Institute of Physics, Goethe-University Frankfurt, 60438 Frankfurt am Main, Germany}

\maketitle

\newpage

\section{Additional experimental results}
\begin{figure*}[h]
    \includegraphics[width=\textwidth]{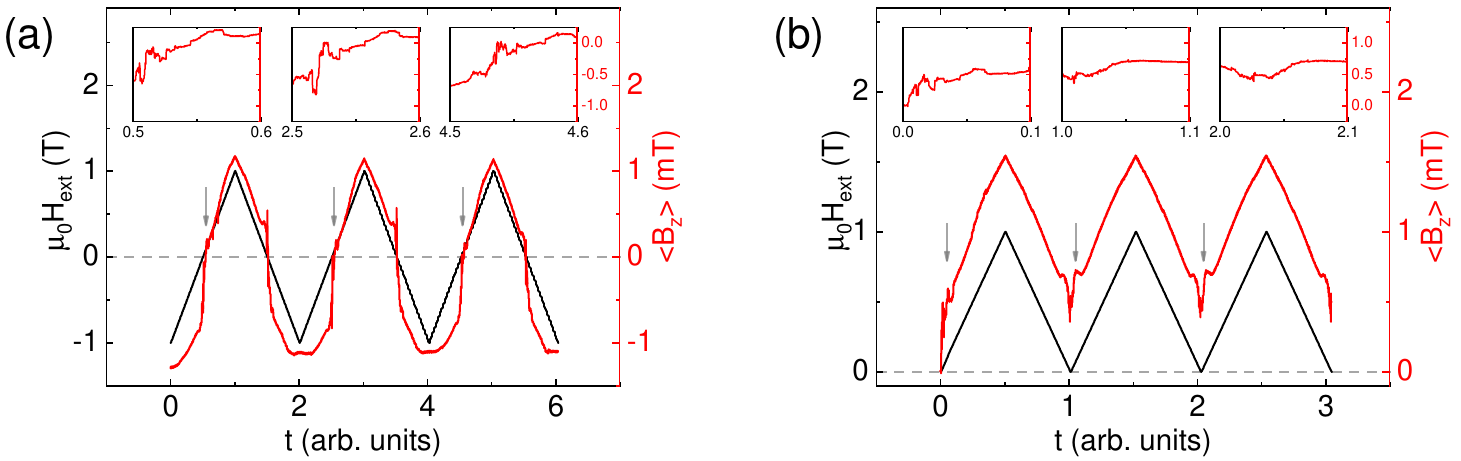}
    \caption{\label{sweeps} Stray field measurements (``plus'' array, $\theta = 95^\circ$) of three consecutive field sweeps (a) in the range $[-1, 1]\,$T and (b) in the range $[0,1]\,$T. The external field is shown in black and the measured stray fields in red. The missing zero-field crossings in (b) result in fewer spikes and steps,  consistent with our model.}
\end{figure*}

\begin{figure*}
  \includegraphics[width=16 cm]{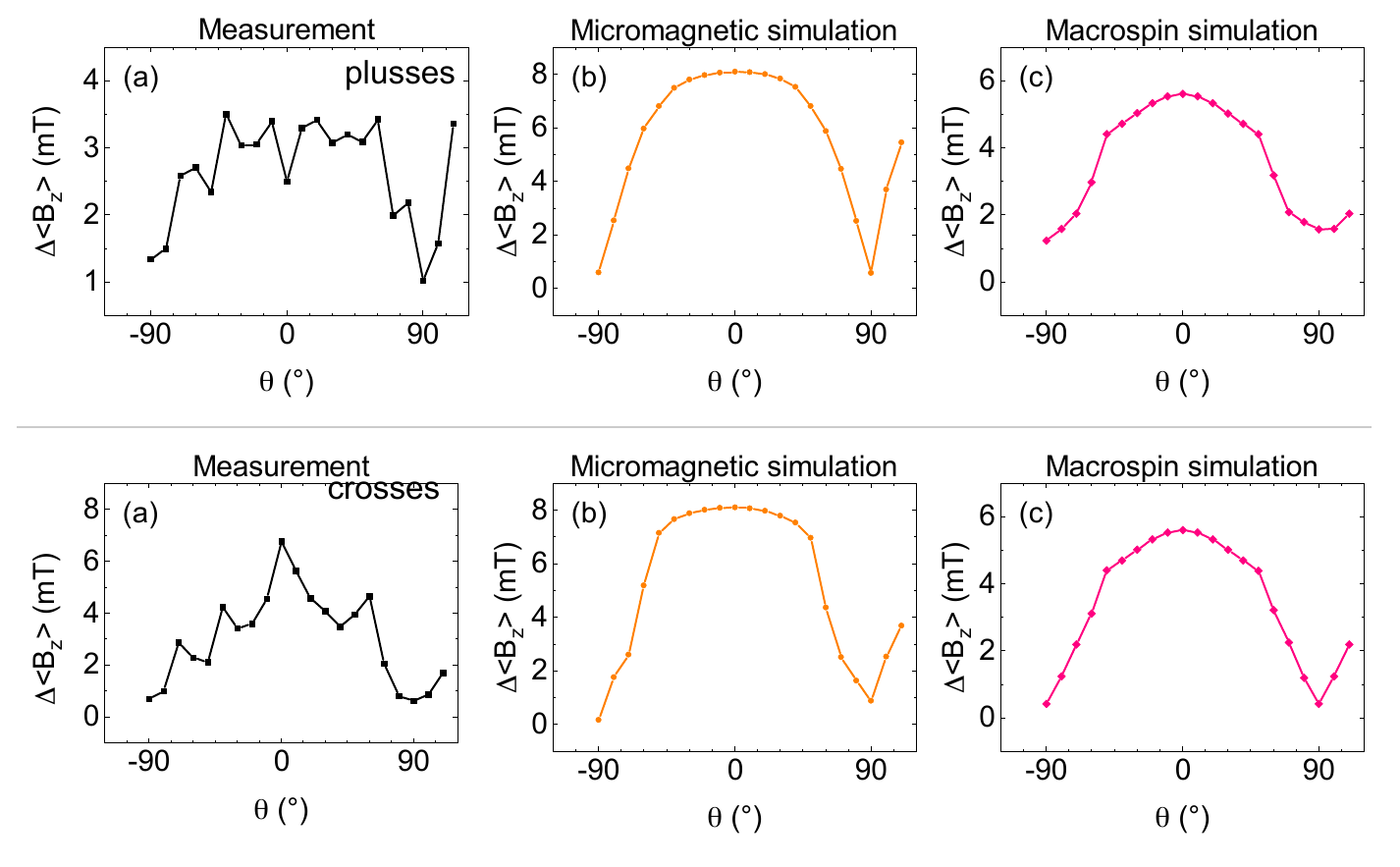}
	\caption{\label{CrDiffBz} Maximum stray field amplitude of a full hysteresis loop $\Delta B_z(\theta)$ as a function of field angle $\theta$ for the plus (top) and cross (bottom) arrays. In the experiment as well as in our simulations we find the minimum value at $\pm$ 90\degree\ whereas the maximum value is located at 0\degree.} 
\end{figure*}

In Fig.~\ref{sweeps}(a) and (b) we show results for three consecutive field sweeps in the range $[-1, 1]\,$T and $[0,1]\,$T, respectively.    
One can clearly see that avoiding zero-field crossing while sweeping the field leads to fewer and smaller spikes and steps as expected from our model. Furthermore, after the first up-sweep one finds almost identical patterns repeating during each field sweep. However, the first up-sweep shows significant differences in both cases. This is due to the fact that the initial orientation of the magnetic moment in each grain is randomly distributed. Only after reaching saturation for the first time the pillars are in a well-defined magnetization state. Note that in our experimental data we find the onset of magnetization switching starting slightly before crossing $\mu_0H_{\mathrm{ext}}=0\,$T. We attribute this to more complex reorientation processes, including competing interactions that are beyond our macrospin model considerations here. However, this accounts only for a small fraction of the switching events and may belong to the vertex central region where the pillars meet and the picture of coarse-grained macrospins fails. 
As a further proof of our coarse-graining model one should find a certain $\theta$-dependence of the total stray field amplitude $\Delta B_z(\theta)  = \langle B_z^\mathrm{max} \rangle - \langle B_z^\mathrm{min} \rangle$ for each hysteresis loop. Importantly, for $\theta = \pm 90$\degree\ the external field is exactly parallel to the Hall cross, i.e.~points along the $x$-direction. At any finite field lowering from saturation the magnetic grains' magnetization is more or less aligned in the $x$-direction as well. Thus, their stray field contribution to $\langle B_z \rangle$ is very small for most fields, except for the fields $H_{\mathrm{C},\phi}$ at which the grains flip. The hysteresis curves at and around $\theta = \pm 90$\degree\ therefore show the most noticeable ``fingerprint'' of the grain switching cascade, however resulting in the smallest total stray field amplitude $\Delta B_z(\theta)$. On decreasing $\theta$, the $z$-component of the grains' magnetization increases and thus $\Delta B_z(\theta)$ increases as well. As a consequence, the relative stray field contribution of the grain switching cascade to $\Delta B_z(\theta)$ becomes smaller, leading to less pronounced spikes and steps with the least contribution at $\theta = 0$\degree, see Fig.~\ref{CrDiffBz}.

\newpage 

\section{Stray field hysteresis loops at additional angles for the "cross" configuration}
Here, we show experimental stray field hysteresis loops as well as results form micromagnetic and macro-spin simulations for additional angles of the "cross" configuration.

\begin{figure*}[h!]
	\includegraphics[width=\textwidth]{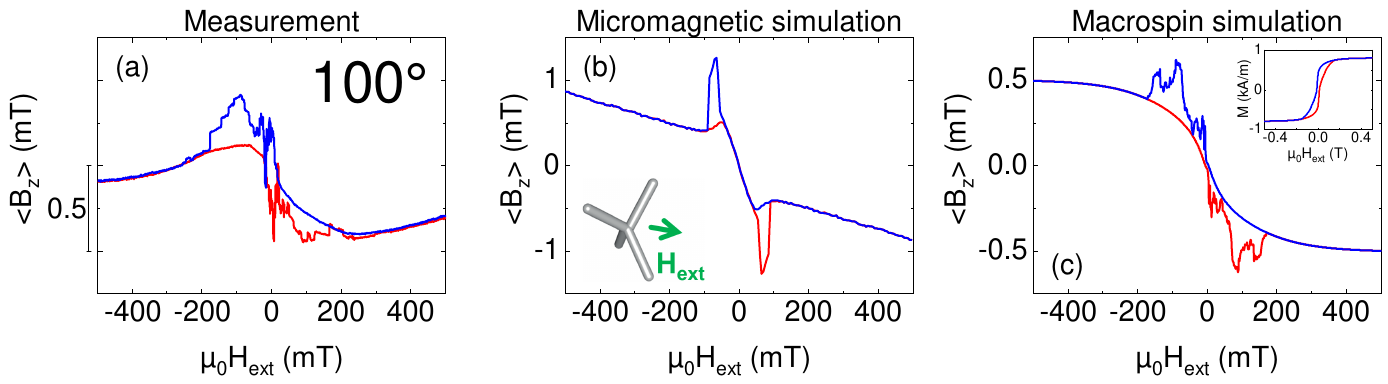}
	\caption{\label{ex_cross_mm_ms_100deg} (a) Measured strayfield hysteresis curves of the "cross" configuration at $\theta =\ang{100}$, (b) micromagnetic and (c) macrospin simulations. The inset of (c) shows the simulated magnetization curves for comparison. The up and down sweeps are coloured in red and blue, respectively. } 
\end{figure*}

\begin{figure*}[h!]
	\includegraphics[width=\textwidth]{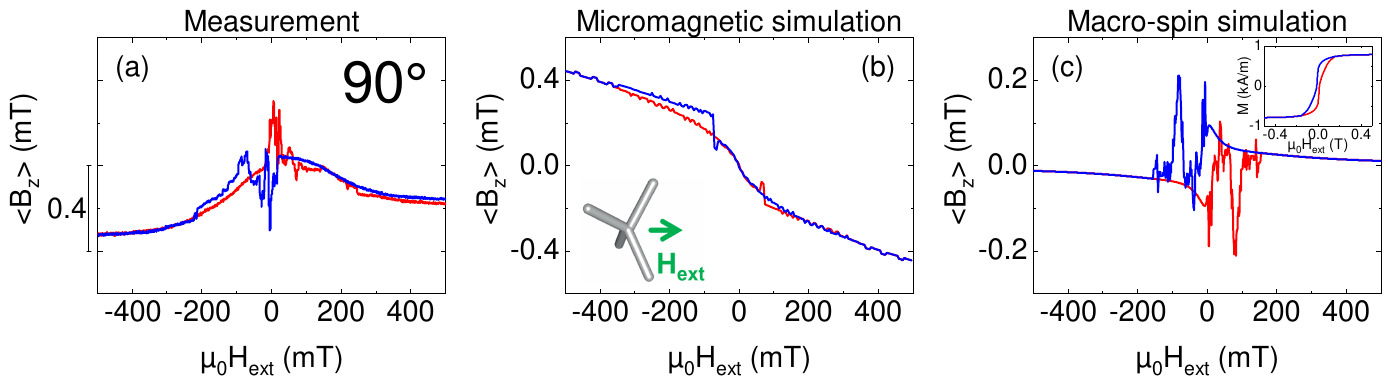}
	\caption{\label{ex_cross_mm_ms_90deg} (a) Measured strayfield hysteresis curves of the "cross" configuration at $\theta =\ang{90}$, (b) micromagnetic and (c) macrospin simulations. The inset of (c) shows the simulated magnetization curves for comparison. The up and down sweeps are coloured in red and blue, respectively.} 
\end{figure*}

\begin{figure*}[h!]
	\includegraphics[width=\textwidth]{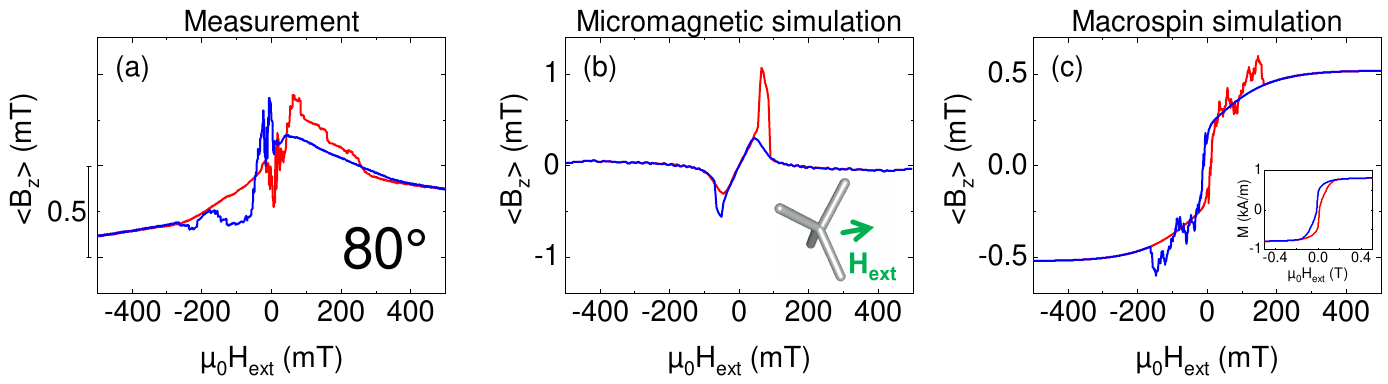}
	\caption{\label{ex_cross_mm_ms_80deg} (a) Measured strayfield hysteresis curves of the "cross" configuration at $\theta =\ang{80}$, (b) micromagnetic and (c) macrospin simulations. The inset of (c) shows the simulated magnetization curves for comparison. The up and down sweeps are coloured in red and blue, respectively.} 
\end{figure*}

\begin{figure*}[h!]
	\includegraphics[width=\textwidth]{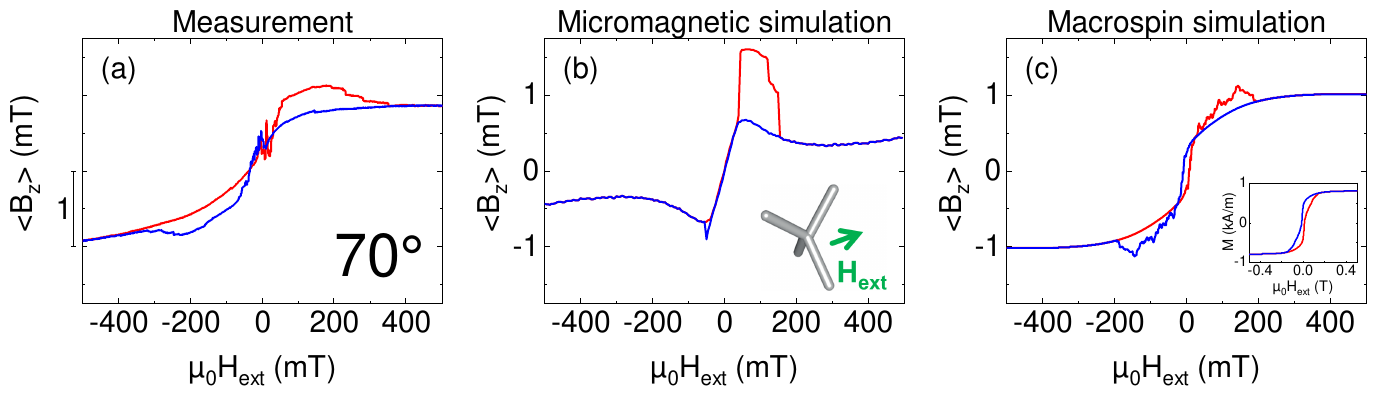}
	\caption{\label{ex_cross_mm_ms_70deg} (a) Measured strayfield hysteresis curves of the "cross" configuration at $\theta =\ang{70}$, (b) micromagnetic and (c) macrospin simulations. The inset of (c) shows the simulated magnetization curves for comparison. The up and down sweeps are coloured in red and blue, respectively.}  
\end{figure*}

\begin{figure*}[h!]
	\includegraphics[width=\textwidth]{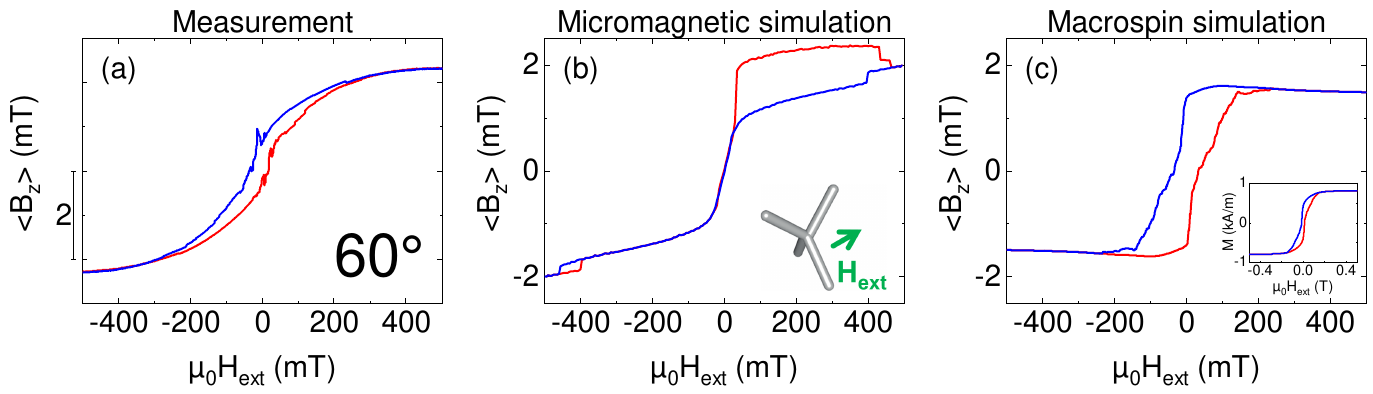}
	\caption{\label{ex_cross_mm_ms_60deg} (a) Measured strayfield hysteresis curves of the "cross" configuration at $\theta =\ang{60}$, (b) micromagnetic and (c) macrospin simulations. The inset of (c) shows the simulated magnetization curves for comparison. The up and down sweeps are coloured in red and blue, respectively.} 
\end{figure*}

\begin{figure*}[h!]
	\includegraphics[width=\textwidth]{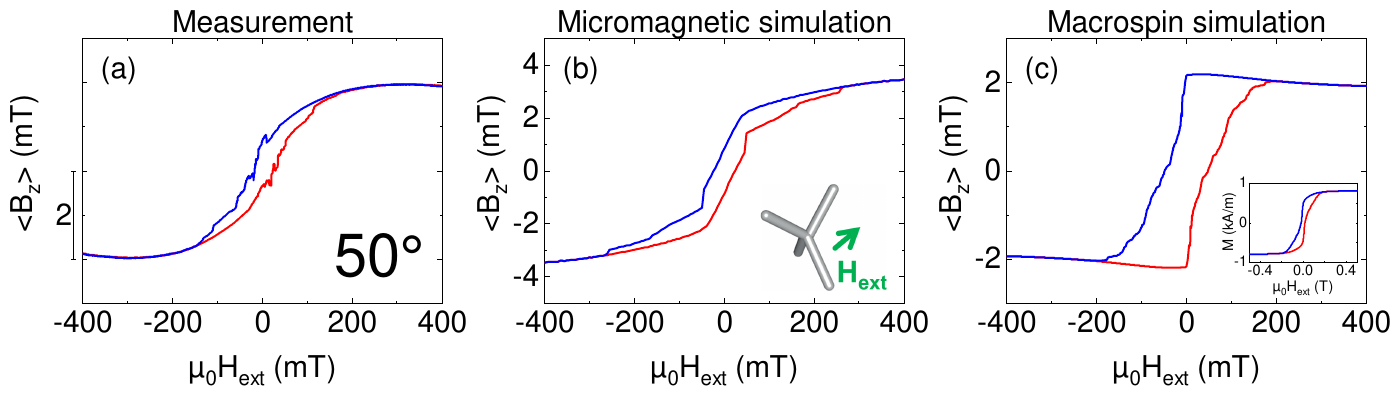}
	\caption{\label{ex_cross_mm_ms_50deg} (a) Measured strayfield hysteresis curves of the "cross" configuration at $\theta =\ang{50}$, (b) micromagnetic and (c) macrospin simulations. The inset of (c) shows the simulated magnetization curves for comparison. The up and down sweeps are coloured in red and blue, respectively.} 
\end{figure*}

\begin{figure*}[h!]
	\includegraphics[width=\textwidth]{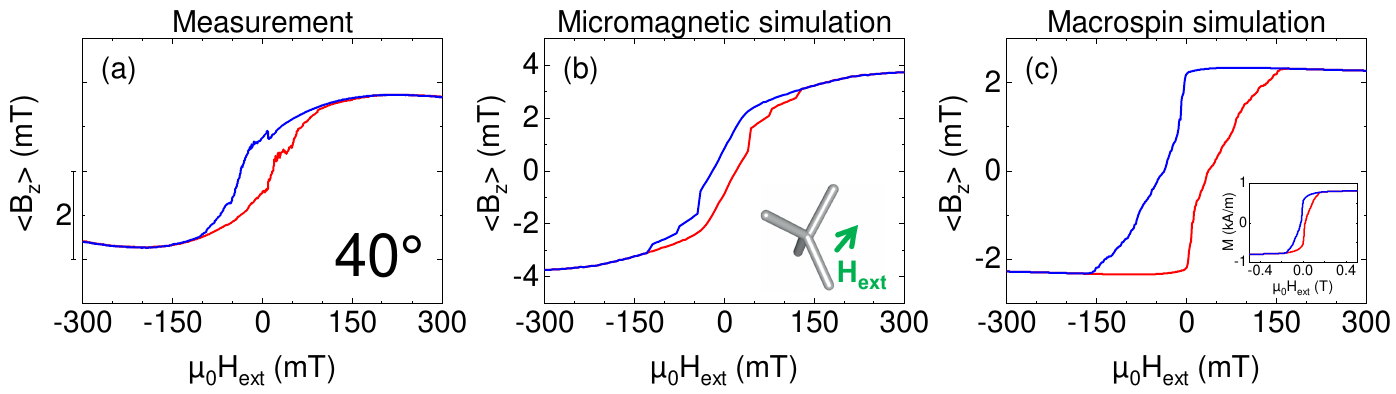}
	\caption{\label{ex_cross_mm_ms_40deg} (a) Measured strayfield hysteresis curves of the "cross" configuration at $\theta =\ang{40}$, (b) micromagnetic and (c) macrospin simulations. The inset of (c) shows the simulated magnetization curves for comparison. The up and down sweeps are coloured in red and blue, respectively.} 
\end{figure*}

\begin{figure*}[h!]
	\includegraphics[width=\textwidth]{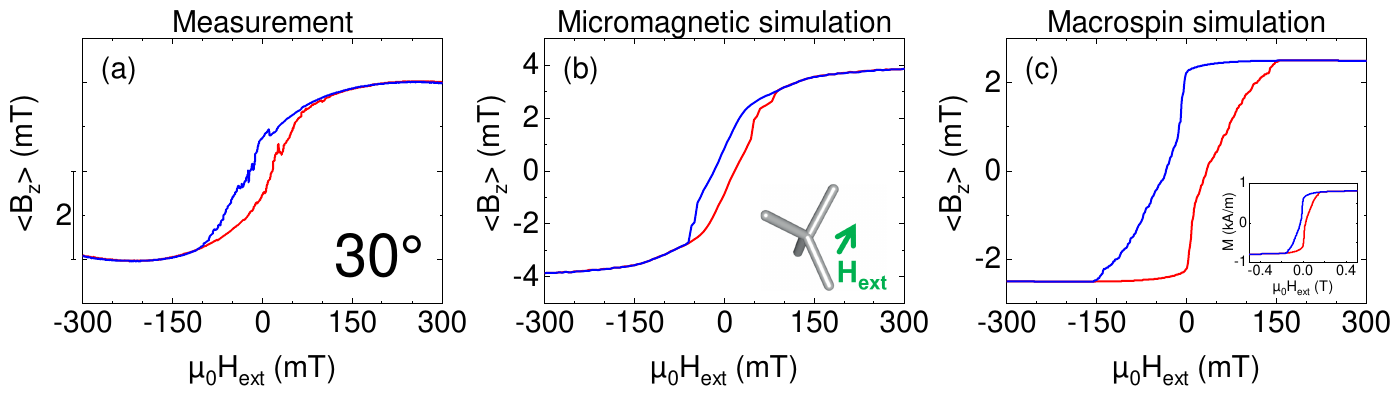}
	\caption{\label{ex_cross_mm_ms_30deg} (a) Measured strayfield hysteresis curves of the "cross" configuration at $\theta =\ang{30}$, (b) micromagnetic and (c) macrospin simulations. The inset of (c) shows the simulated magnetization curves for comparison. The up and down sweeps are coloured in red and blue, respectively.}
\end{figure*}

\begin{figure*}[h!]
	\includegraphics[width=\textwidth]{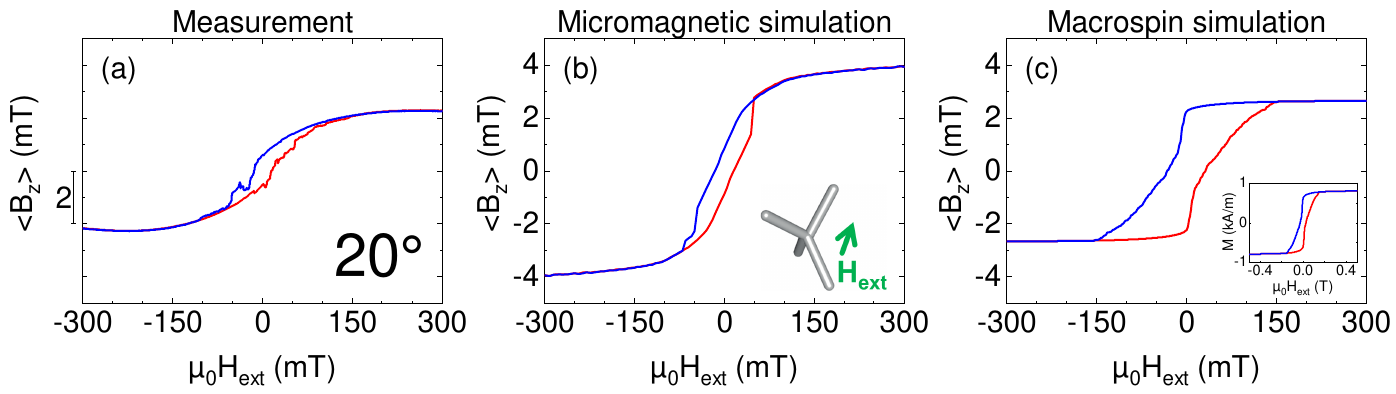}
	\caption{\label{ex_cross_mm_ms_20deg} (a) Measured strayfield hysteresis curves of the "cross" configuration at $\theta =\ang{20}$, (b) micromagnetic and (c) macrospin simulations. The inset of (c) shows the simulated magnetization curves for comparison. The up and down sweeps are coloured in red and blue, respectively.} 
\end{figure*}

\begin{figure*}[h!]
	\includegraphics[width=\textwidth]{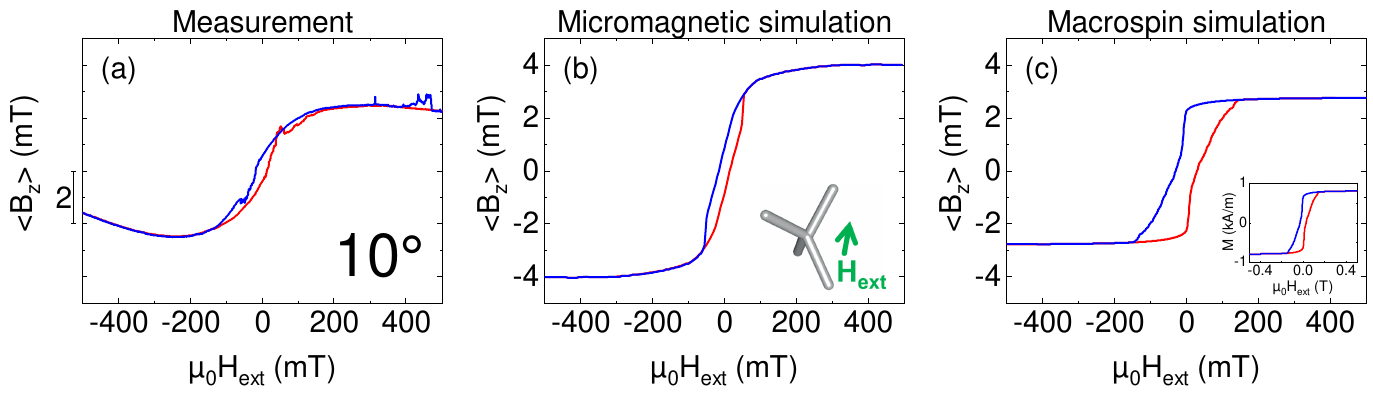}
	\caption{\label{ex_cross_mm_ms_0deg} (a) Measured strayfield hysteresis curves of the "cross" configuration at $\theta =\ang{10}$, (b) micromagnetic and (c) macrospin simulations. The inset of (c) shows the simulated magnetization curves for comparison. The up and down sweeps are coloured in red and blue, respectively.} 
\end{figure*}

\begin{figure*}[h!]
	\includegraphics[width=\textwidth]{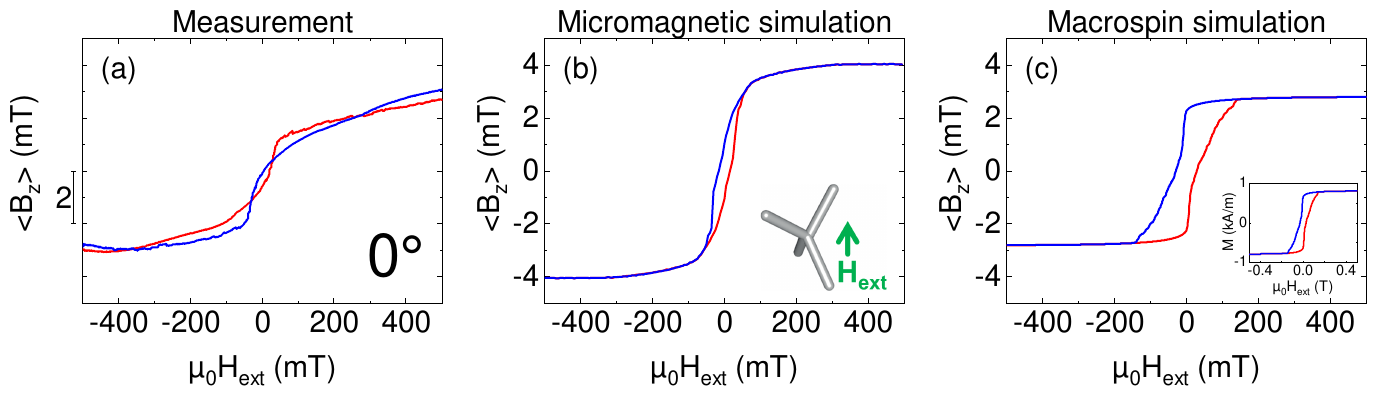}
	\caption{\label{ex_cross_mm_ms_0deg} (a) Measured strayfield hysteresis curves of the "cross" configuration at $\theta =\ang{0}$, (b) micromagnetic and (c) macrospin simulations. The inset of (c) shows the simulated magnetization curves for comparison. The up and down sweeps are coloured in red and blue, respectively.} 
\end{figure*}

\begin{figure*}[h!]
	\includegraphics[width=\textwidth]{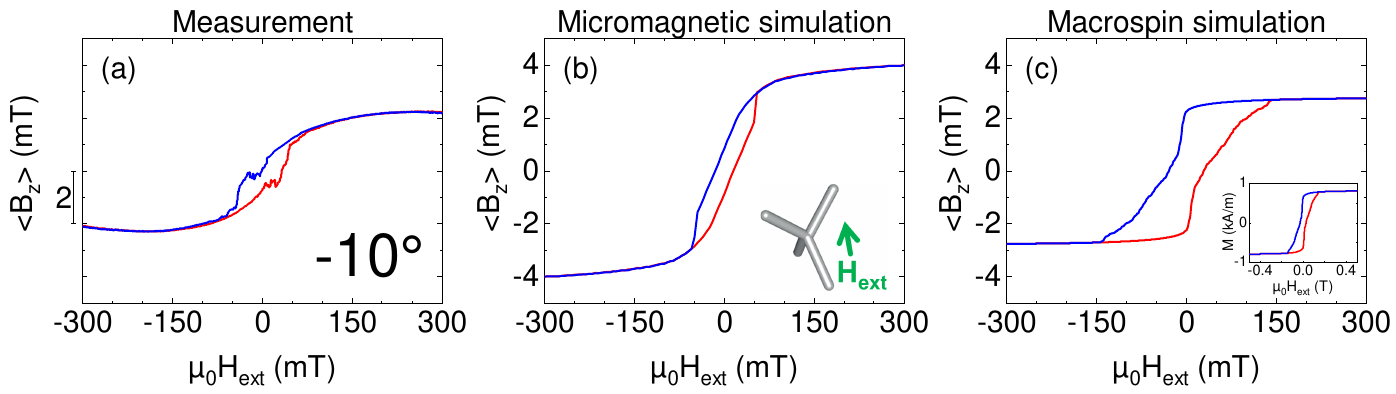}
	\caption{\label{ex_cross_mm_ms_-10deg} (a) Measured strayfield hysteresis curves of the "cross" configuration at $\theta =\ang{-10}$, (b) micromagnetic and (c) macrospin simulations. The inset of (c) shows the simulated magnetization curves for comparison. The up and down sweeps are coloured in red and blue, respectively.}  
\end{figure*}

\begin{figure*}[h!]
	\includegraphics[width=\textwidth]{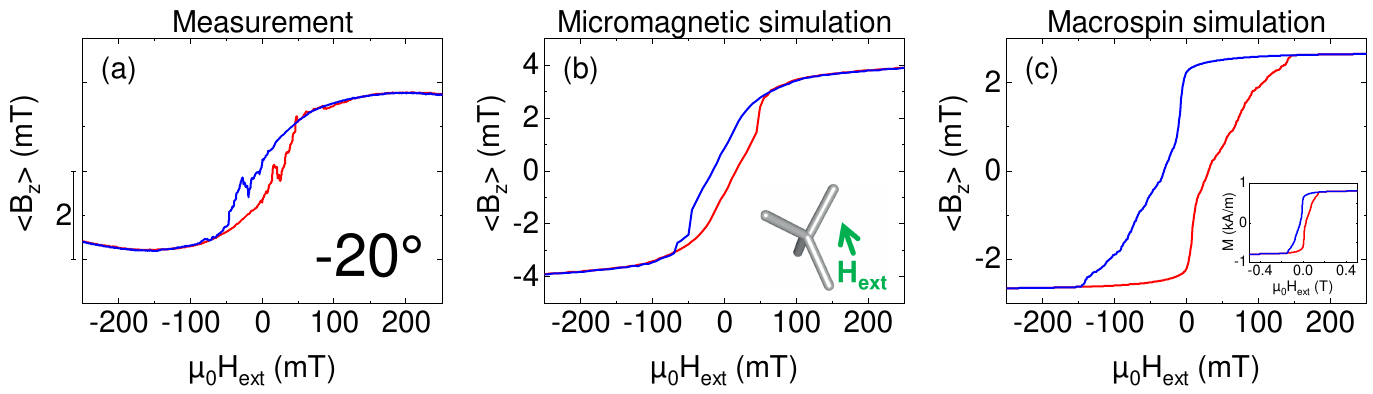}
	\caption{\label{ex_cross_mm_ms_-20deg} (a) Measured strayfield hysteresis curves of the "cross" configuration at $\theta =\ang{-20}$, (b) micromagnetic and (c) macrospin simulations. The inset of (c) shows the simulated magnetization curves for comparison. The up and down sweeps are coloured in red and blue, respectively.} 
\end{figure*}

\begin{figure*}[h!]
	\includegraphics[width=\textwidth]{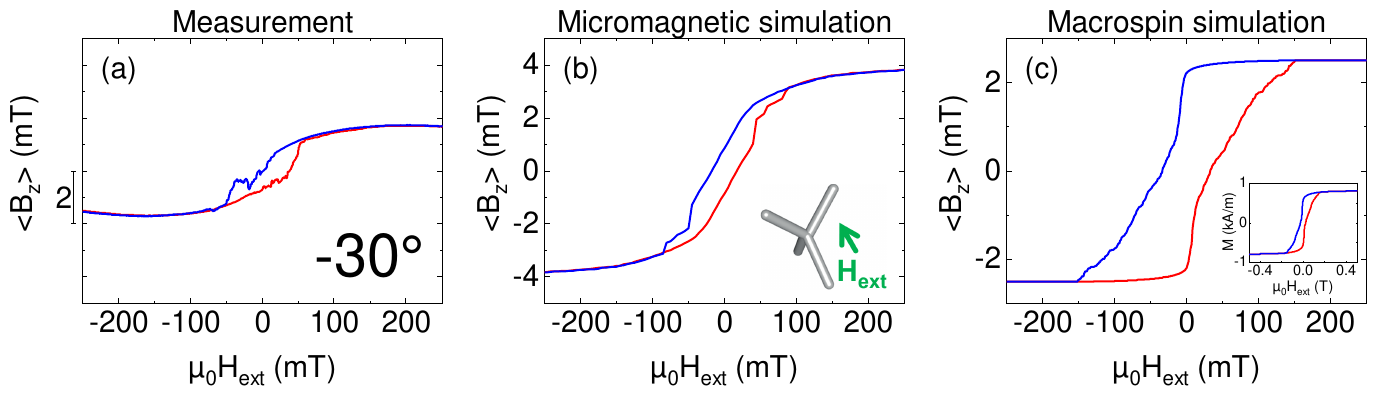}
	\caption{\label{ex_cross_mm_ms_-30deg} (a) Measured strayfield hysteresis curves of the "cross" configuration at $\theta =\ang{-30}$, (b) micromagnetic and (c) macrospin simulations. The inset of (c) shows the simulated magnetization curves for comparison. The up and down sweeps are coloured in red and blue, respectively.} 
\end{figure*}

\begin{figure*}[h!]
	\includegraphics[width=\textwidth]{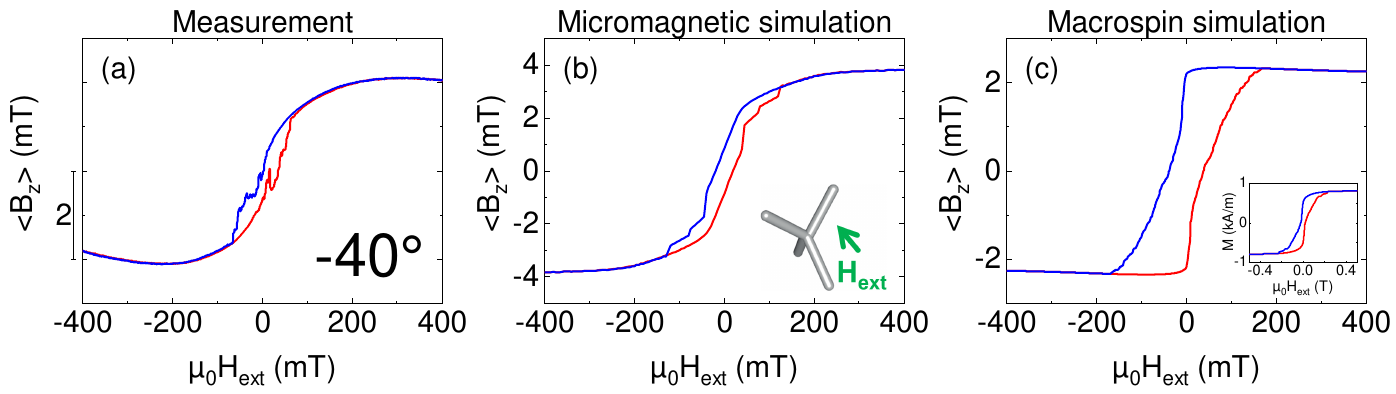}
	\caption{\label{ex_cross_mm_ms_-40deg} (a) Measured strayfield hysteresis curves of the "cross" configuration at $\theta =\ang{-40}$, (b) micromagnetic and (c) macrospin simulations. The inset of (c) shows the simulated magnetization curves for comparison. The up and down sweeps are coloured in red and blue, respectively.} 
\end{figure*}

\begin{figure*}[h!]
	\includegraphics[width=\textwidth]{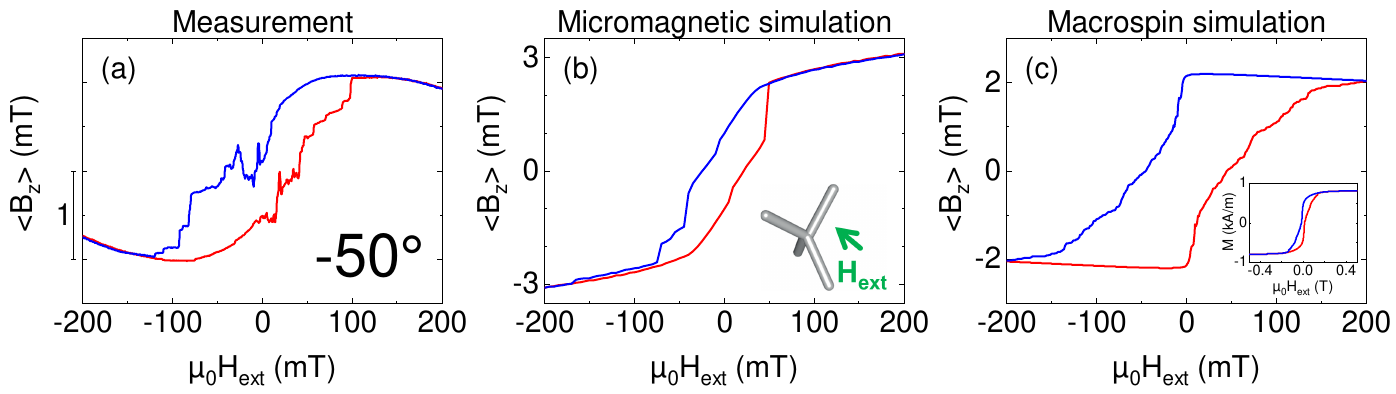}
	\caption{\label{ex_cross_mm_ms_-50deg} (a) Measured strayfield hysteresis curves of the "cross" configuration at $\theta =\ang{-50}$, (b) micromagnetic and (c) macrospin simulations. The inset of (c) shows the simulated magnetization curves for comparison. The up and down sweeps are coloured in red and blue, respectively.} 
\end{figure*}

\begin{figure*}[h!]
	\includegraphics[width=\textwidth]{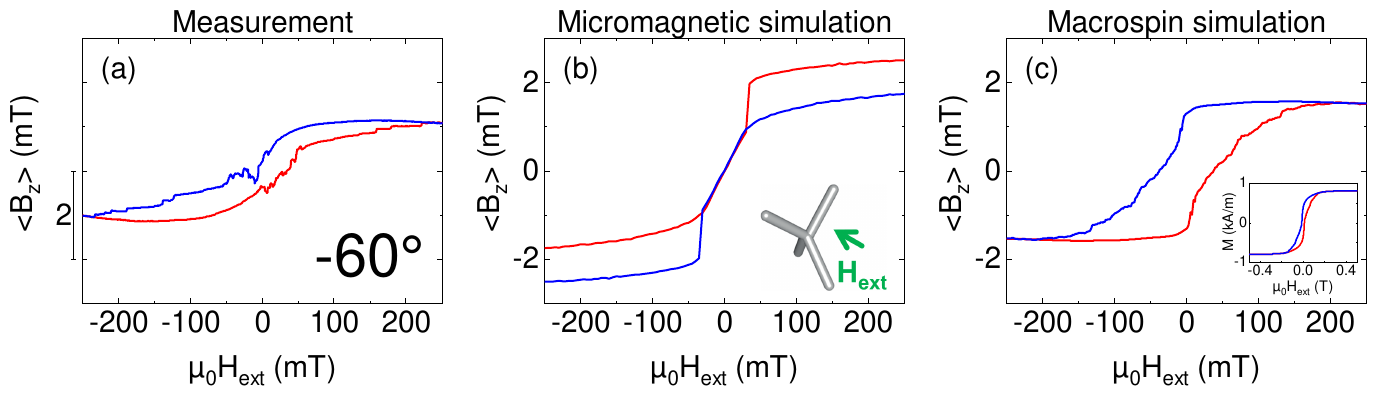}
	\caption{\label{ex_cross_mm_ms_-60deg} (a) Measured strayfield hysteresis curves of the "cross" configuration at $\theta =\ang{-60}$, (b) micromagnetic and (c) macrospin simulations. The inset of (c) shows the simulated magnetization curves for comparison. The up and down sweeps are coloured in red and blue, respectively.} 
\end{figure*}

\begin{figure*}[h!]
	\includegraphics[width=\textwidth]{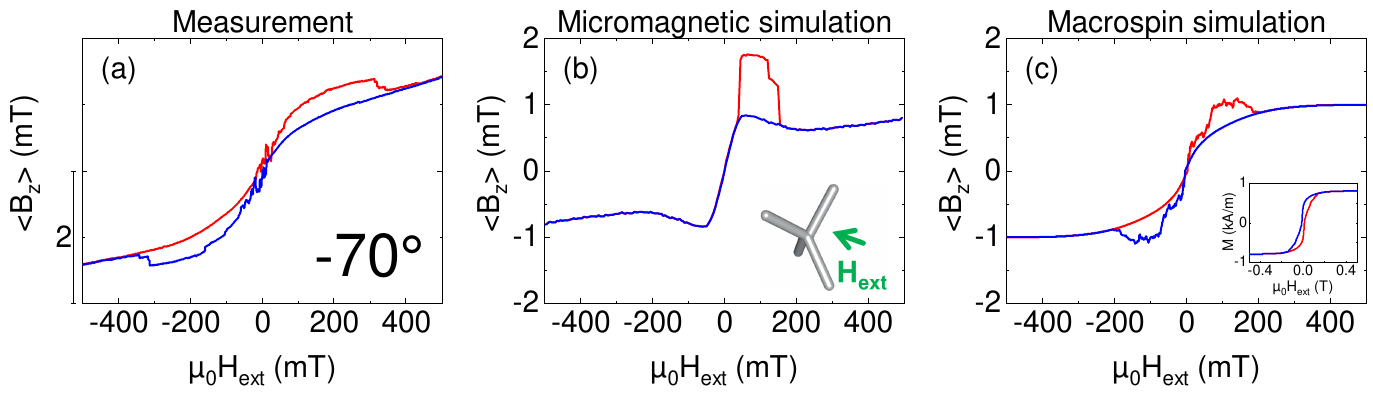}
	\caption{\label{ex_cross_mm_ms_-70deg} (a) Measured strayfield hysteresis curves of the "cross" configuration at $\theta =\ang{-70}$, (b) micromagnetic and (c) macrospin simulations. The inset of (c) shows the simulated magnetization curves for comparison. The up and down sweeps are coloured in red and blue, respectively.} 
\end{figure*}

\begin{figure*}[h!]
	\includegraphics[width=\textwidth]{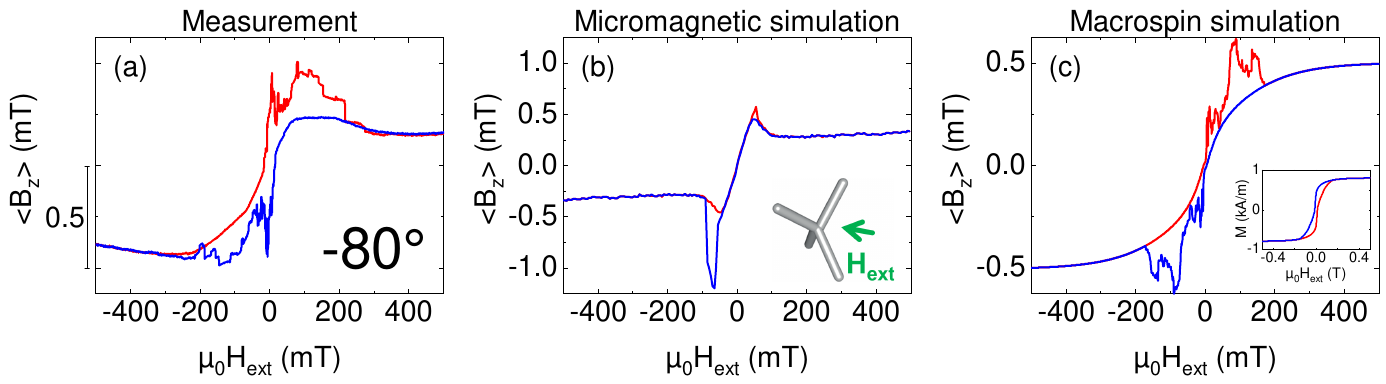}
	\caption{\label{ex_cross_mm_ms_-80deg} (a) Measured strayfield hysteresis curves of the "cross" configuration at $\theta =\ang{-80}$, (b) micromagnetic and (c) macrospin simulations. The inset of (c) shows the simulated magnetization curves for comparison. The up and down sweeps are coloured in red and blue, respectively.} 
\end{figure*}

\begin{figure*}[h!]
	\includegraphics[width=\textwidth]{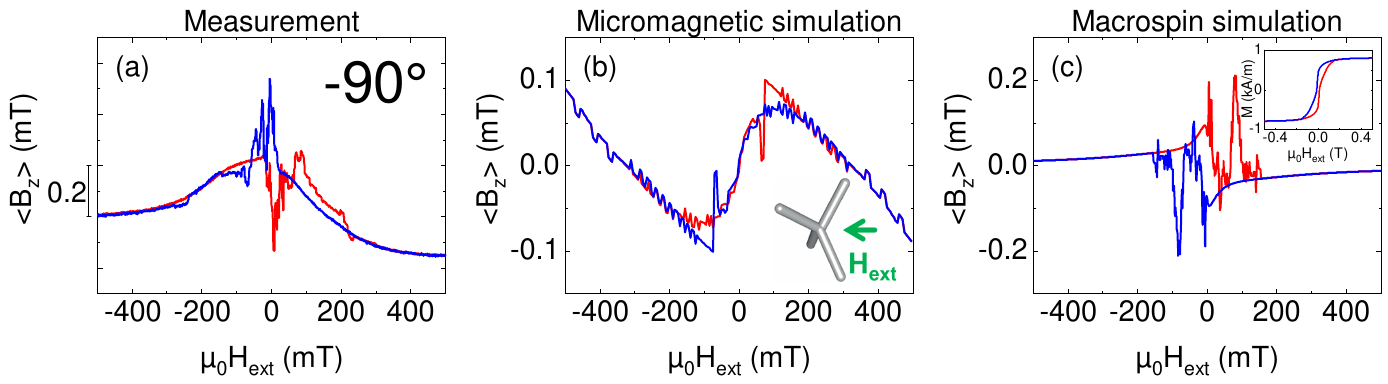}
	\caption{\label{ex_mm_ms_-90deg} (a) Measured strayfield hysteresis curves of the "cross" configuration at $\theta =\ang{-90}$, (b) micromagnetic and (c) macrospin simulations. The inset of (c) shows the simulated magnetization curves for comparison. The up and down sweeps are coloured in red and blue, respectively.} 
\end{figure*}

\newpage
\clearpage
\section{Stray field hysteresis loops at additional angles for the "plus" configuration}
Here, we show experimental stray field hysteresis loops as well as results form micromagnetic and macro-spin simulations for additional angles of the "plus" configuration.

\begin{figure*}[h!]
	\includegraphics[width=\textwidth]{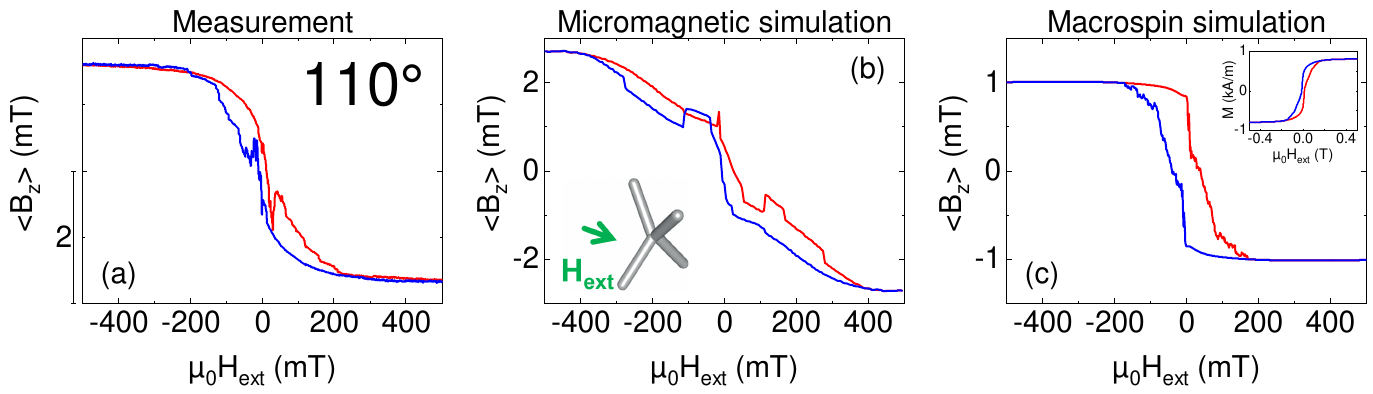}
	\caption{\label{ex_plus_mm_ms_110deg} (a) Measured strayfield hysteresis curves of the "plus" configuration at $\theta =\ang{110}$, (b) micromagnetic and (c) macrospin simulations. The inset of (c) shows the simulated magnetization curves for comparison. The up and down sweeps are coloured in red and blue, respectively.} 
\end{figure*}

\begin{figure*}[h!]
	\includegraphics[width=\textwidth]{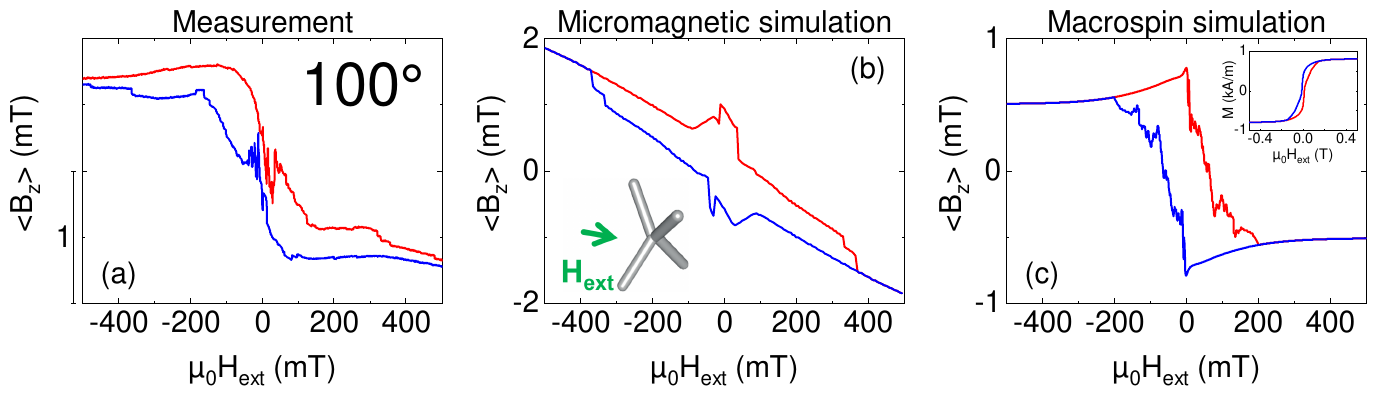}
	\caption{\label{ex_plus_mm_ms_100deg} (a) Measured strayfield hysteresis curves of the "plus" configuration at $\theta =\ang{100}$, (b) micromagnetic and (c) macrospin simulations. The inset of (c) shows the simulated magnetization curves for comparison. The up and down sweeps are coloured in red and blue, respectively.} 
\end{figure*}

\begin{figure*}[h!]
	\includegraphics[width=\textwidth]{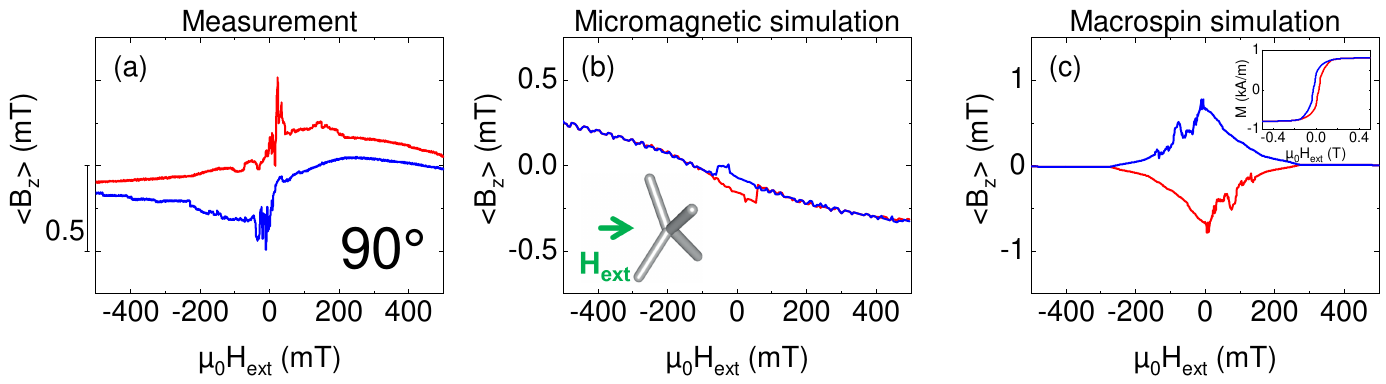}
	\caption{\label{ex_plus_mm_ms_90deg} (a) Measured strayfield hysteresis curves of the "plus" configuration at $\theta =\ang{90}$, (b) micromagnetic and (c) macrospin simulations. The inset of (c) shows the simulated magnetization curves for comparison. The up and down sweeps are coloured in red and blue, respectively.} 
\end{figure*}

\begin{figure*}[h!]
	\includegraphics[width=\textwidth]{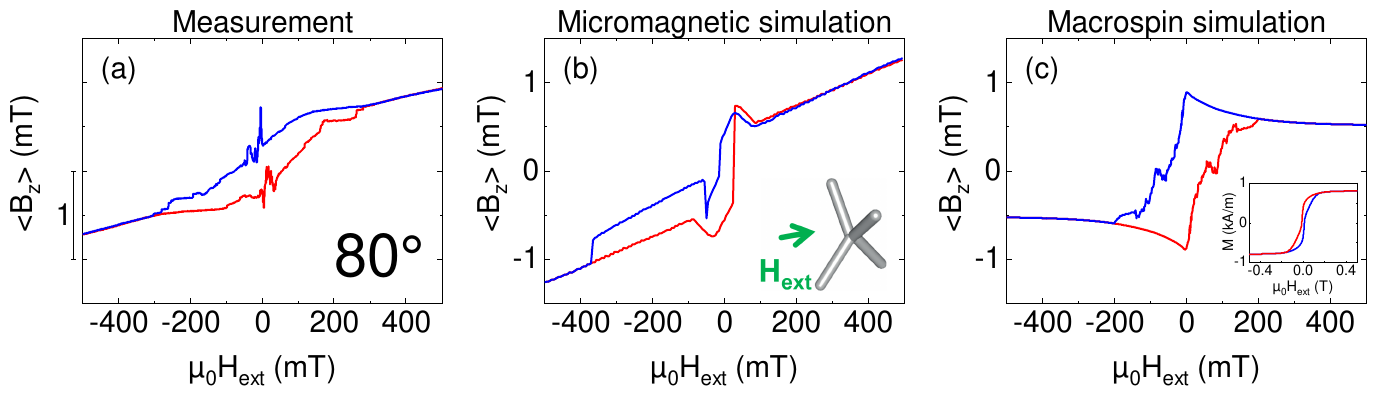}
	\caption{\label{ex_plus_mm_ms_80deg} (a) Measured strayfield hysteresis curves of the "plus" configuration at $\theta =\ang{80}$, (b) micromagnetic and (c) macrospin simulations. The inset of (c) shows the simulated magnetization curves for comparison. The up and down sweeps are coloured in red and blue, respectively.} 
\end{figure*}

\begin{figure*}[h!]
	\includegraphics[width=\textwidth]{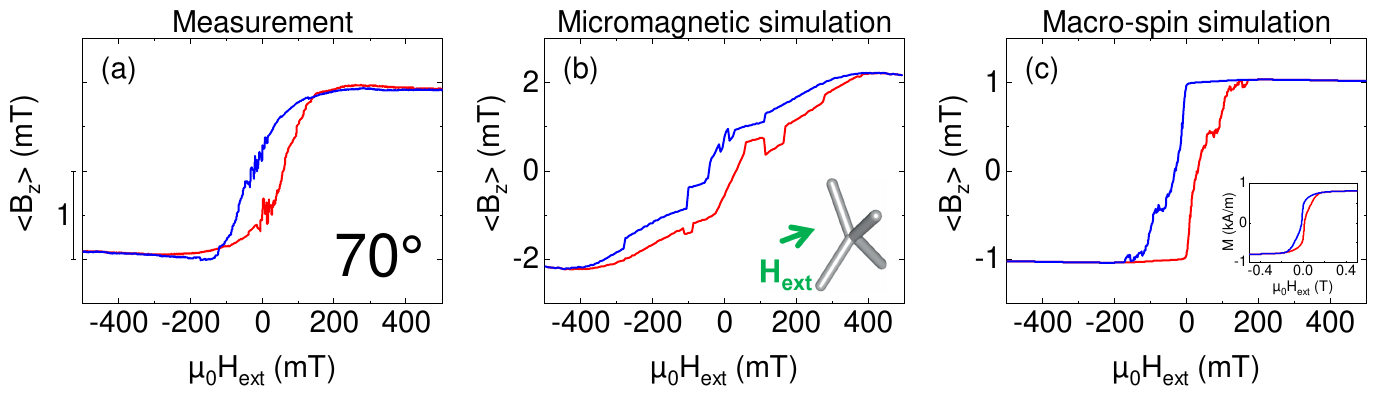}
	\caption{\label{ex_plus_mm_ms_70deg} (a) Measured strayfield hysteresis curves of the "plus" configuration at $\theta =\ang{70}$, (b) micromagnetic and (c) macrospin simulations. The inset of (c) shows the simulated magnetization curves for comparison. The up and down sweeps are coloured in red and blue, respectively.}  
\end{figure*}

\begin{figure*}[h!]
	\includegraphics[width=\textwidth]{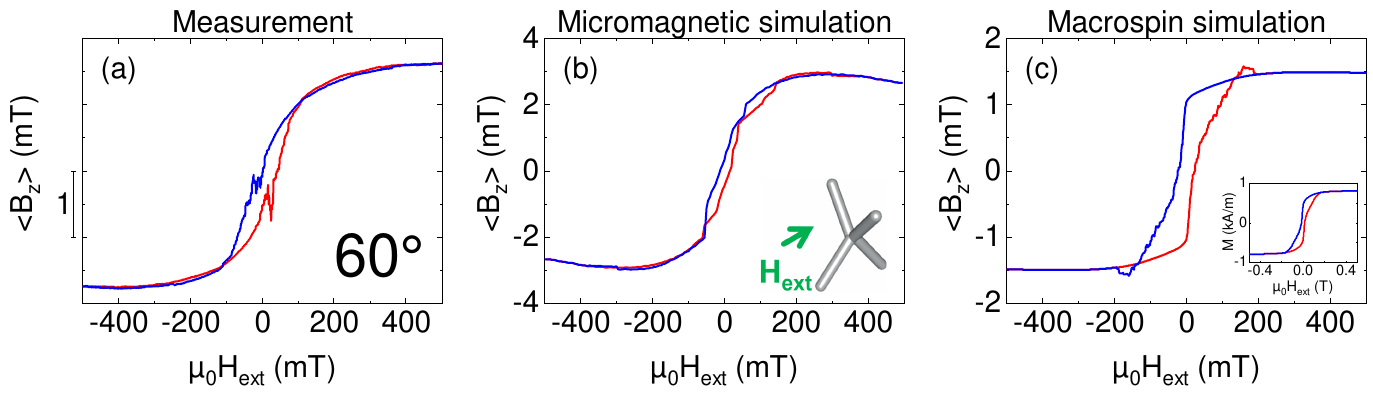}
	\caption{\label{ex_plus_mm_ms_60deg} (a) Measured strayfield hysteresis curves of the "plus" configuration at $\theta =\ang{60}$, (b) micromagnetic and (c) macrospin simulations. The inset of (c) shows the simulated magnetization curves for comparison. The up and down sweeps are coloured in red and blue, respectively.} 
\end{figure*}

\begin{figure*}[h!]
	\includegraphics[width=\textwidth]{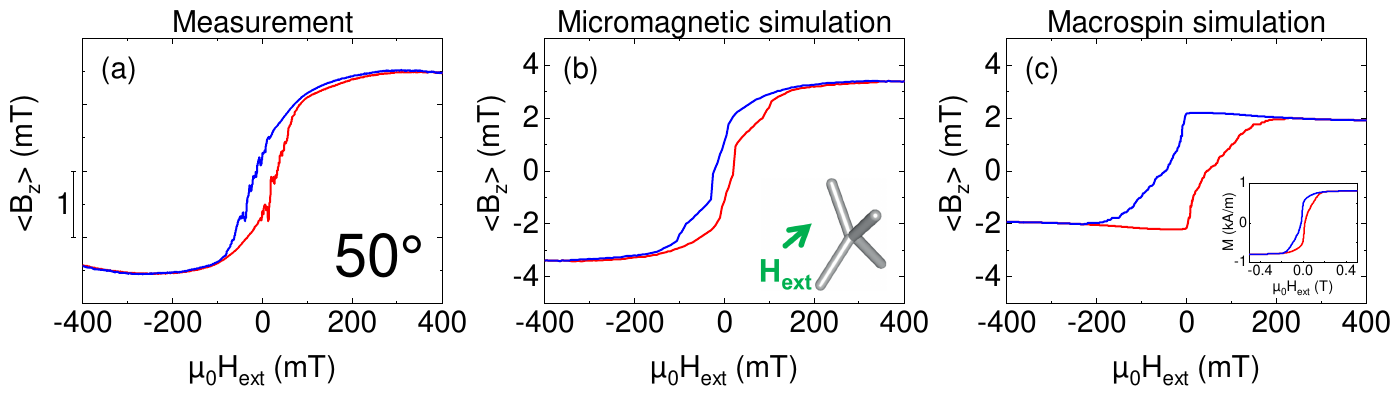}
	\caption{\label{ex_plus_mm_ms_50deg} (a) Measured strayfield hysteresis curves of the "plus" configuration at $\theta =\ang{50}$, (b) micromagnetic and (c) macrospin simulations. The inset of (c) shows the simulated magnetization curves for comparison. The up and down sweeps are coloured in red and blue, respectively.} 
\end{figure*}

\begin{figure*}[h!]
	\includegraphics[width=\textwidth]{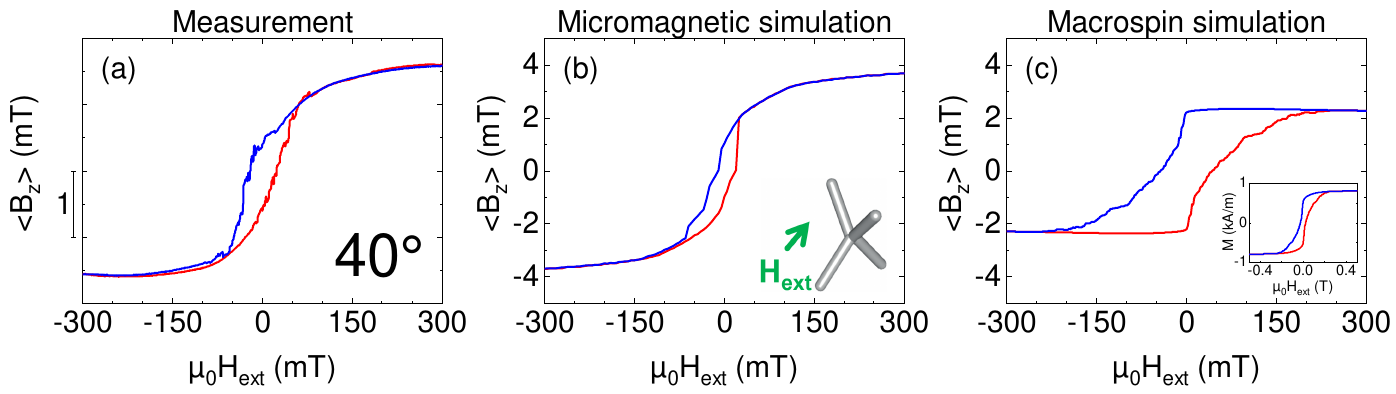}
	\caption{\label{ex_plus_mm_ms_40deg} (a) Measured strayfield hysteresis curves of the "plus" configuration at $\theta =\ang{40}$, (b) micromagnetic and (c) macrospin simulations. The inset of (c) shows the simulated magnetization curves for comparison. The up and down sweeps are coloured in red and blue, respectively.} 
\end{figure*}

\begin{figure*}[h!]
	\includegraphics[width=\textwidth]{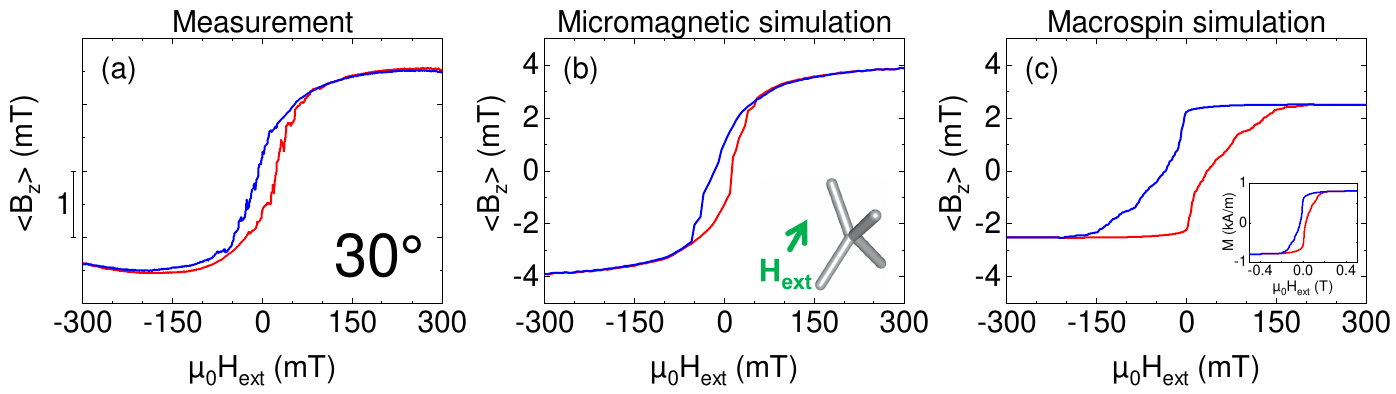}
	\caption{\label{ex_plus_mm_ms_30deg} (a) Measured strayfield hysteresis curves of the "plus" configuration at $\theta =\ang{30}$, (b) micromagnetic and (c) macrospin simulations. The inset of (c) shows the simulated magnetization curves for comparison. The up and down sweeps are coloured in red and blue, respectively.}
\end{figure*}

\begin{figure*}[h!]
	\includegraphics[width=\textwidth]{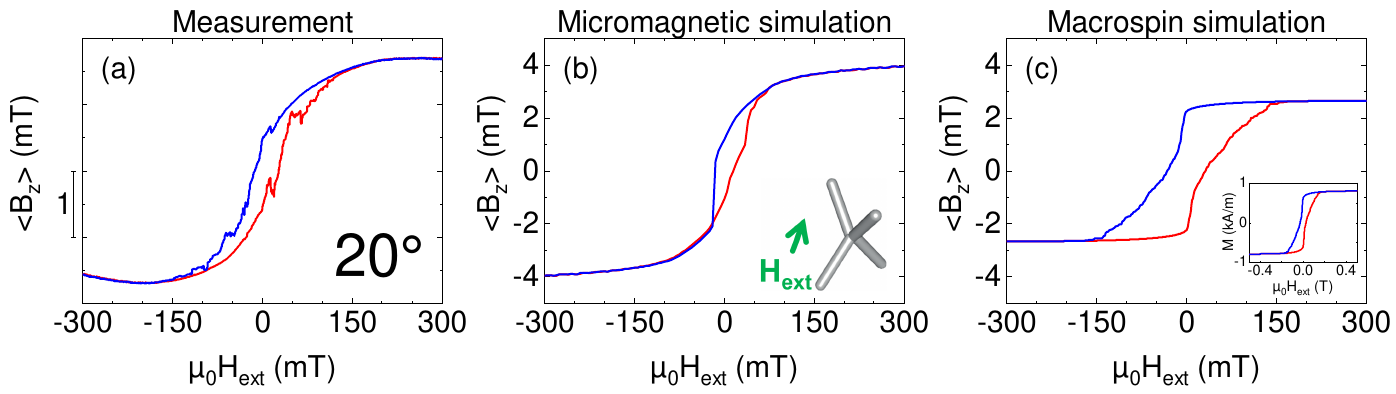}
	\caption{\label{ex_plus_mm_ms_20deg} (a) Measured strayfield hysteresis curves of the "plus" configuration at $\theta =\ang{20}$, (b) micromagnetic and (c) macrospin simulations. The inset of (c) shows the simulated magnetization curves for comparison. The up and down sweeps are coloured in red and blue, respectively.} 
\end{figure*}

\begin{figure*}[h!]
	\includegraphics[width=\textwidth]{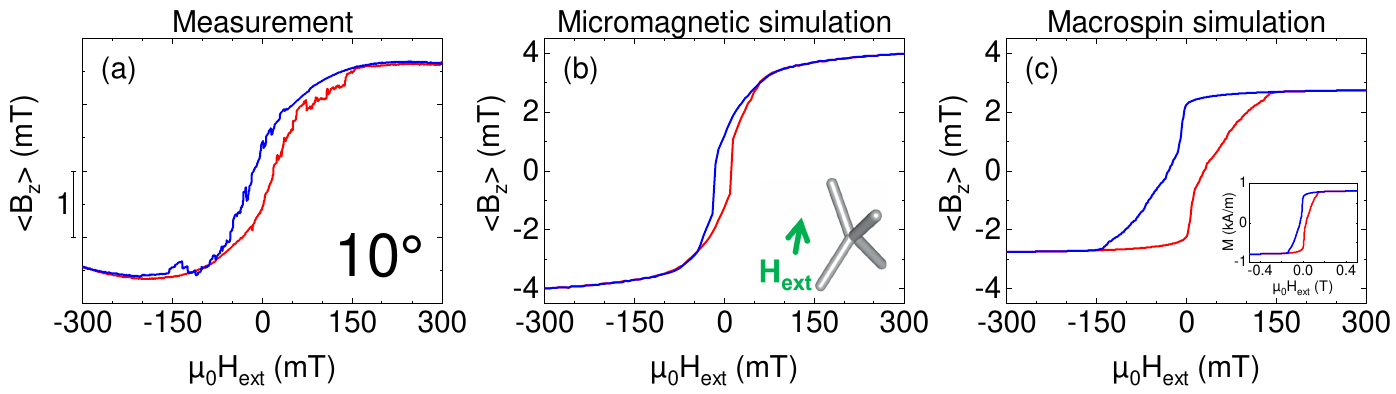}
	\caption{\label{ex_plus_mm_ms_0deg} (a) Measured strayfield hysteresis curves of the "plus" configuration at $\theta =\ang{10}$, (b) micromagnetic and (c) macrospin simulations. The inset of (c) shows the simulated magnetization curves for comparison. The up and down sweeps are coloured in red and blue, respectively.} 
\end{figure*}

\begin{figure*}[h!]
	\includegraphics[width=\textwidth]{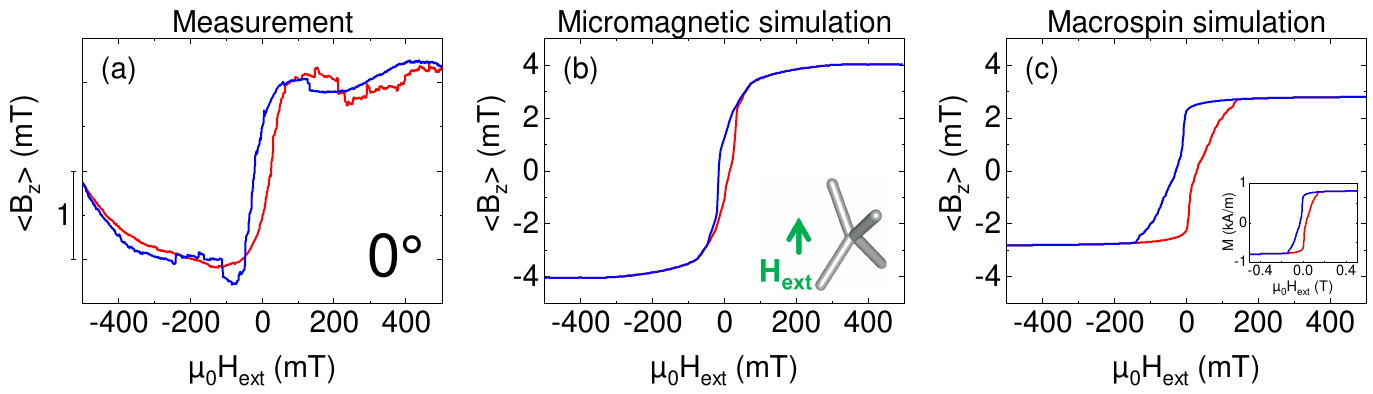}
	\caption{\label{ex_plus_mm_ms_0deg} (a) Measured strayfield hysteresis curves of the "plus" configuration at $\theta =\ang{0}$, (b) micromagnetic and (c) macrospin simulations. The inset of (c) shows the simulated magnetization curves for comparison. The up and down sweeps are coloured in red and blue, respectively.} 
\end{figure*}

\begin{figure*}[h!]
	\includegraphics[width=\textwidth]{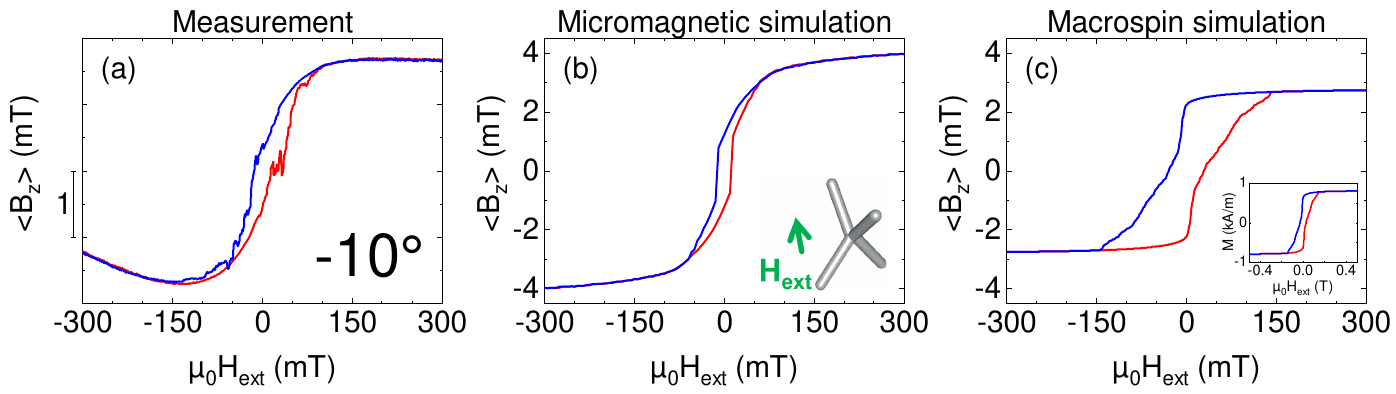}
	\caption{\label{ex_plus_mm_ms_-10deg} (a) Measured strayfield hysteresis curves of the "plus" configuration at $\theta =\ang{-10}$, (b) micromagnetic and (c) macrospin simulations. The inset of (c) shows the simulated magnetization curves for comparison. The up and down sweeps are coloured in red and blue, respectively.}  
\end{figure*}

\begin{figure*}[h!]
	\includegraphics[width=\textwidth]{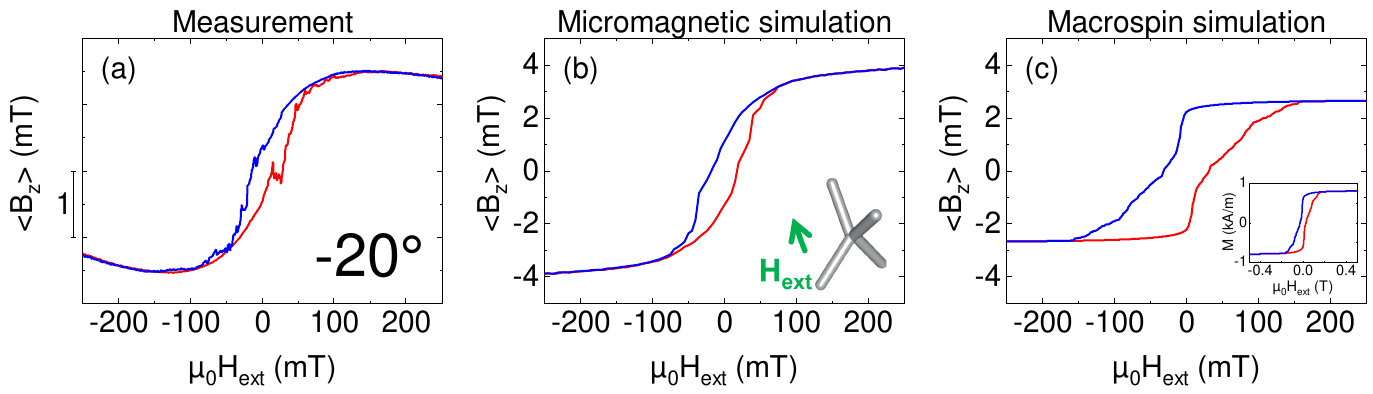}
	\caption{\label{ex_plus_mm_ms_-20deg} (a) Measured strayfield hysteresis curves of the "plus" configuration at $\theta =\ang{-20}$, (b) micromagnetic and (c) macrospin simulations. The inset of (c) shows the simulated magnetization curves for comparison. The up and down sweeps are coloured in red and blue, respectively.} 
\end{figure*}

\begin{figure*}[h!]
	\includegraphics[width=\textwidth]{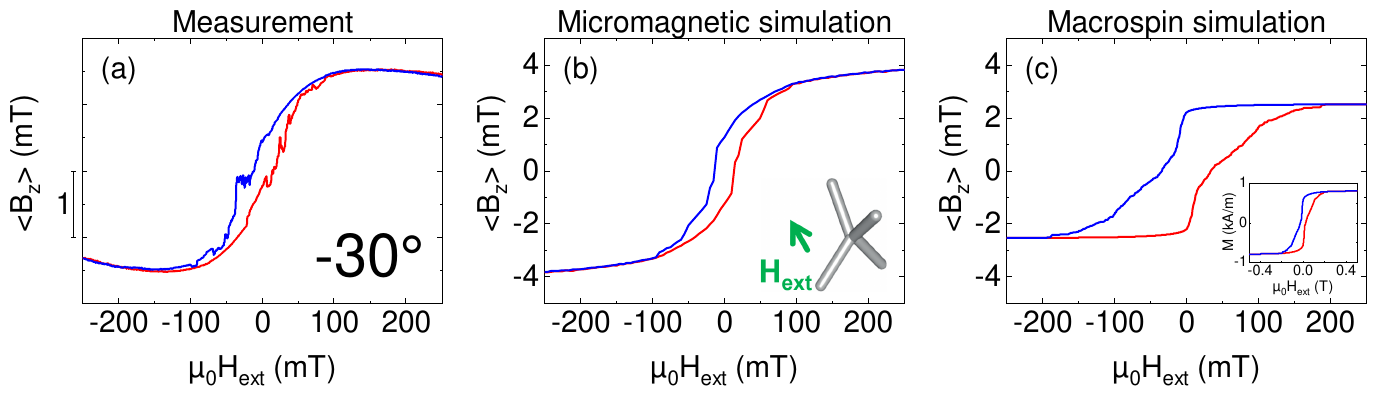}
	\caption{\label{ex_plus_mm_ms_-30deg} (a) Measured strayfield hysteresis curves of the "plus" configuration at $\theta =\ang{-30}$, (b) micromagnetic and (c) macrospin simulations. The inset of (c) shows the simulated magnetization curves for comparison. The up and down sweeps are coloured in red and blue, respectively.} 
\end{figure*}

\begin{figure*}[h!]
	\includegraphics[width=\textwidth]{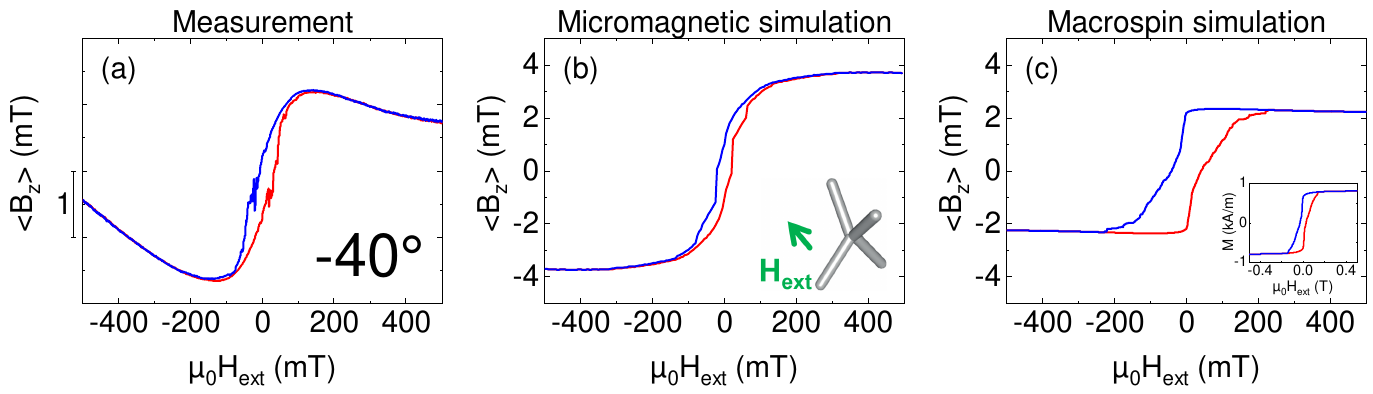}
	\caption{\label{ex_plus_mm_ms_-40deg} (a) Measured strayfield hysteresis curves of the "plus" configuration at $\theta =\ang{-40}$, (b) micromagnetic and (c) macrospin simulations. The inset of (c) shows the simulated magnetization curves for comparison. The up and down sweeps are coloured in red and blue, respectively.} 
\end{figure*}

\begin{figure*}[h!]
	\includegraphics[width=\textwidth]{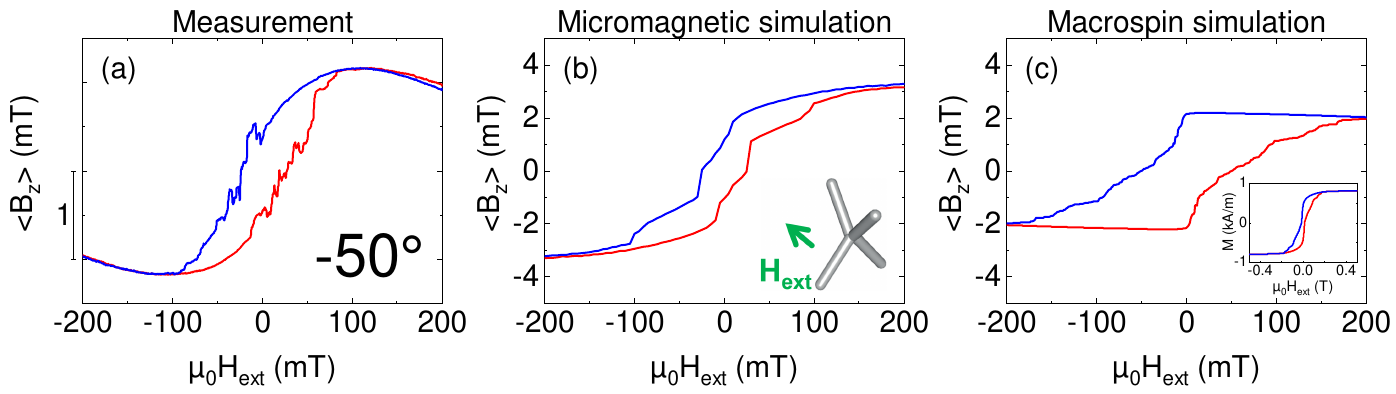}
	\caption{\label{ex_plus_mm_ms_-50deg} (a) Measured strayfield hysteresis curves of the "plus" configuration at $\theta =\ang{-50}$, (b) micromagnetic and (c) macrospin simulations. The inset of (c) shows the simulated magnetization curves for comparison. The up and down sweeps are coloured in red and blue, respectively.} 
\end{figure*}

\begin{figure*}[h!]
	\includegraphics[width=\textwidth]{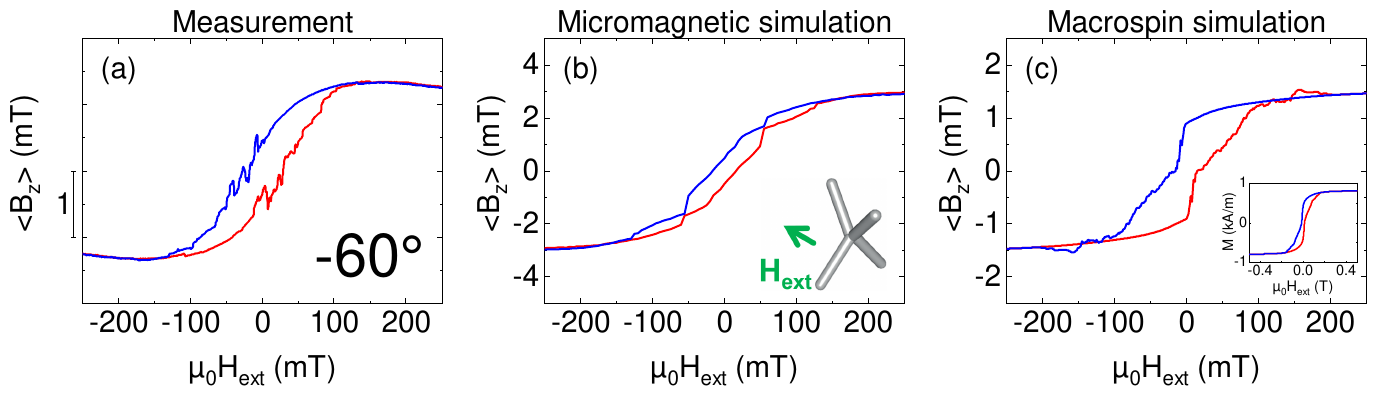}
	\caption{\label{ex_plus_mm_ms_-60deg} (a) Measured strayfield hysteresis curves of the "plus" configuration at $\theta =\ang{-60}$, (b) micromagnetic and (c) macrospin simulations. The inset of (c) shows the simulated magnetization curves for comparison. The up and down sweeps are coloured in red and blue, respectively.} 
\end{figure*}

\begin{figure*}[h!]
	\includegraphics[width=\textwidth]{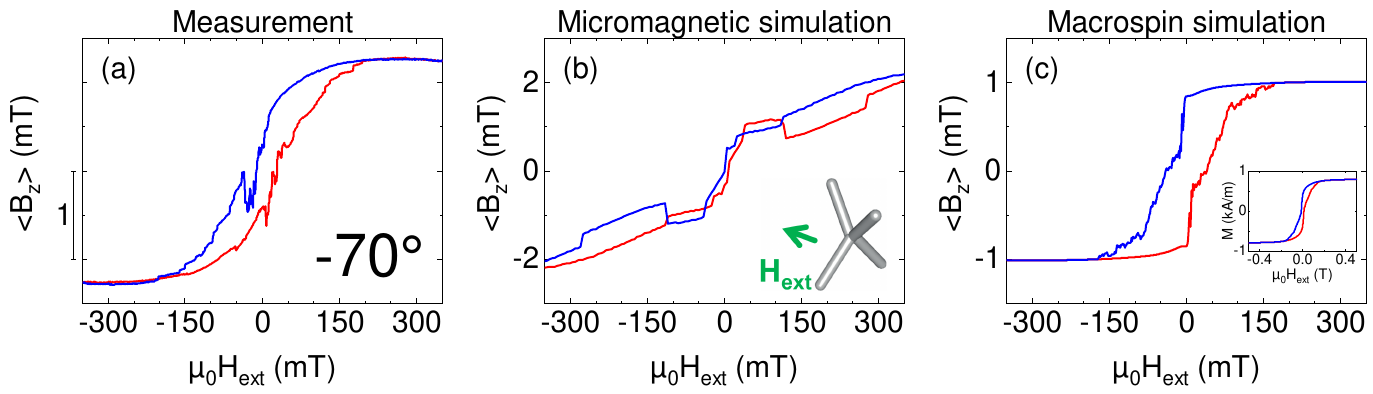}
	\caption{\label{ex_plus_mm_ms_-70deg} (a) Measured strayfield hysteresis curves of the "plus" configuration at $\theta =\ang{-70}$, (b) micromagnetic and (c) macrospin simulations. The inset of (c) shows the simulated magnetization curves for comparison. The up and down sweeps are coloured in red and blue, respectively.} 
\end{figure*}

\begin{figure*}[h!]
	\includegraphics[width=\textwidth]{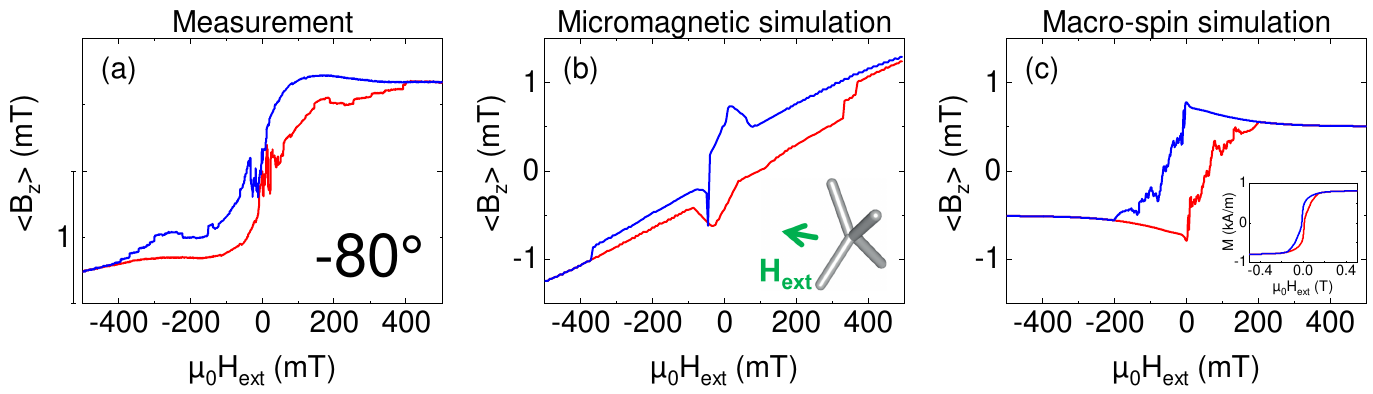}
	\caption{\label{ex_plus_mm_ms_-80deg} (a) Measured strayfield hysteresis curves of the "plus" configuration at $\theta =\ang{-80}$, (b) micromagnetic and (c) macrospin simulations. The inset of (c) shows the simulated magnetization curves for comparison. The up and down sweeps are coloured in red and blue, respectively.} 
\end{figure*}

\begin{figure*}[h!]
	\includegraphics[width=\textwidth]{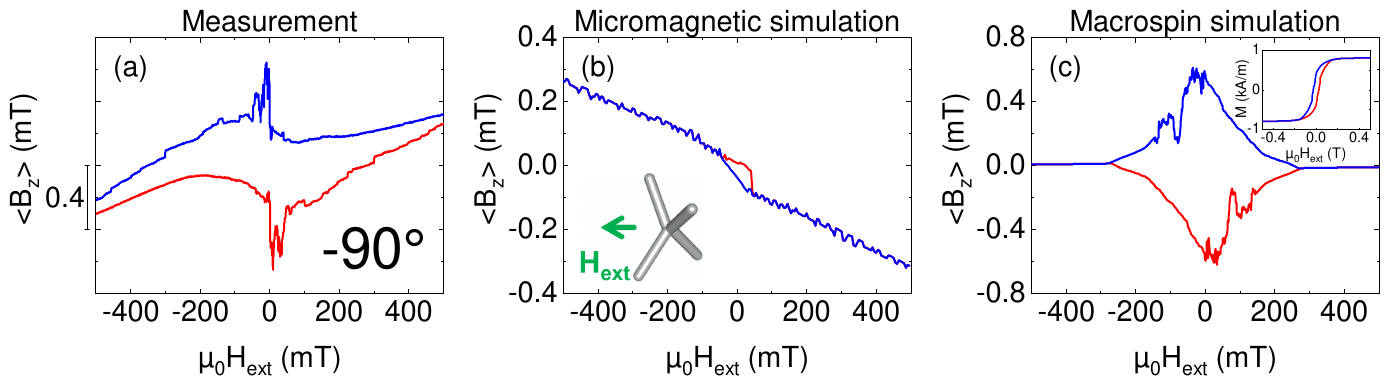}
	\caption{\label{ex_mm_ms_-90deg} (a) Measured strayfield hysteresis curves of the "plus" configuration at $\theta =\ang{-90}$, (b) micromagnetic and (c) macrospin simulations. The inset of (c) shows the simulated magnetization curves for comparison. The up and down sweeps are coloured in red and blue, respectively.} 
\end{figure*}

